\documentclass[11pt,a4paper]{article}
\usepackage{jheppub}
\usepackage{amsthm,bm,mathrsfs}

\newcommand{\be}{\begin{equation}}
\newcommand{\ee}{\end{equation}}
\newcommand{\ba}{\begin{eqnarray}}
\newcommand{\ea}{\end{eqnarray}}

\newcommand{\bra}[1]{\left<\,{#1}\,\right|}
\newcommand{\ket}[1]{\left|\,{#1}\,\right>}
\newcommand{\m}[1]{$\mathop{#1}$}

\newcommand{\la}{\langle}
\newcommand{\ra}{\rangle}

\newcommand{\pd}[2]{\langle\,{#1}\,|\,{#2}\,\rangle}
\newcommand{\ppd}[2]{\langle\!\langle\,{#1}\,|\,{#2}\,\rangle\!\rangle}

\newcommand{\st}{\scriptstyle}
\newcommand{\sst}{\scriptscriptstyle}

\newcommand{\nn}{\nonumber\\}
\newcommand{\e}{&=&}

\newcommand{\xav}[1]{{#1}}

\author[a]{Xavier Bekaert}
\author[b]{Euihun Joung}
\author[c]{Jihad Mourad}

\affiliation[a]{Laboratoire de Math\'ematiques et Physique Th\'eorique\footnote{
Unit\'e Mixte de Recherche $6083$ du CNRS,
F\'ed\'eration de Recherche $2964$ Denis Poisson},\\
Universit\'e Fran\c{c}ois Rabelais, Parc de Grandmount,
37200 Tours, France}
\affiliation[b]{Scuola Normale Superiore and INFN\\
Piazza dei Cavalieri 7, 56126 Pisa, Italy}
\affiliation[c]{AstroParticule et Cosmologie\footnote{
Unit\'e Mixte de Recherche $7164$ du CNRS},\\
Universit\'e Paris VII, B\^atiment Condorcet, 75205 Paris Cedex 13, France}

\emailAdd{bekaert@lmpt.univ-tours.fr}
\emailAdd{euihun.joung@sns.it}
\emailAdd{mourad@apc.univ-paris7.fr}

\title{Effective action in a higher-spin background}

\abstract{
We consider a free massless scalar field coupled to an infinite
tower of background higher-spin gauge fields via minimal coupling to the traceless conserved currents.
The set of Abelian gauge transformations is deformed to the non-Abelian group  of
unitary operators acting on the scalar field.
The gauge invariant effective action is computed perturbatively in the external
fields. The structure of the various (divergent or finite) terms is determined. In particular, the
quadratic part of the logarithmically divergent (or of the finite) 
term is expressed in terms of curvatures and related to conformal higher-spin gravity.
The generalized  higher-spin Weyl anomalies are also determined. The relation with the theory of
interacting higher-spin gauge fields on anti de Sitter spacetime via the holographic correspondence is
discussed.
}

\begin{document}

\maketitle

\flushbottom

\section{Introduction}

Gauge fields with spins greater than two are not needed for the
description of the present day experiments and observations.
Moreover several no-go theorems suggest that, beyond the free 
level, consistent higher-spin gauge theories are unusual (see
\textit{e.g.} \cite{Bekaert:2010hw} for an introductory review).
On the other hand, the understanding of the theory of higher-spin
gauge fields is expected to give valuable insights into the
symmetries of string theory. Indeed,  string theory contains an
infinite number of massive higher-spin modes and their presence is
responsible to a large extent for the nice ultraviolet properties
of string amplitudes. Hence, the tensionless limit may reveal the
most symmetric phase of string theory (see \textit{e.g.} the early
work \cite{Gross:1988ue} pushing forward this idea). In this
limit, an infinite number of higher-spin gauge fields appear and
the understanding of consistent higher-spin gauge theories may be
very helpful to describe this phase (see \textit{e.g.}
\cite{Sagnotti:2010at
} and refs therein for some
recent works). A classically consistent theory of interacting higher-spin
gauge fields  in (anti) de Sitter spacetime ((A)dS) is known 
(see \textit{e.g.}
\cite{Vasiliev:2004cp
} for some
reviews). A very simple description of this theory was proposed
by Klebanov and Polyakov via the AdS/CFT correspondence
\cite{Klebanov:2002ja} pursuing earlier insights
\cite{Sundborg:2000wp
, Mikhailov:2002bp,Sezgin:2002rt} in this direction. This holographic description
involves $N$ free massless scalar fields on the boundary and the
infinite set of $O(N)$-singlet Noether currents corresponding to
the infinite tower of gauge fields in the bulk. A more precise
statement of this correspondence is that the effective action of
massless scalar fields in the presence of external higher-spin
fields, coupled minimally via the bilinear currents, is given, in
the semi-classical regime,
 by
the on-shell action of interacting higher-spin gauge fields
expressed in terms of the boundary data. In the present case, the
one-loop effective action is actually exact since the path
integral is Gaussian (the scalar fields are free and their
currents are bilinear).\footnote{Through the holographic
dictionary, the translation of this property (somewhat
unremarkable for a \emph{free} theory) becomes rather striking:
the dual interacting higher-spin theory should not receive any
quantum correction! For unbroken higher-spin gauge symmetries,
this is plausible since the group of symmetries may be big enough
to eliminate any non-trivial counter-term (as advocated a while ago
by Fradkin \cite{Fradkin:1990kr}).} This boundary effective action
is the subject of this paper.

From the point of view of the conformal field theory (CFT),
the effective action is usually interpreted as the generating
functional of the connected correlation functions of the Noether
currents, in which case the gauge fields are mere auxiliary
sources. From another point of view (advocated in
\cite{Liu:1998bu
}), the ultraviolet divergent
(perturbatively local) part of the effective action may also be
interpreted as an induced conformal gravity action for the gauge
fields. Higher-spin conformal gravity theories have been
introduced at the quadratic level in the metric-like formulation
by Fradkin and Tseytlin \cite{Fradkin:1985am} in dimension
four and generalized for any dimension by Segal
\cite{Segal:2002gd}. Their relation with the bulk/boundary
correspondence has been investigated by Metsaev (see
\cite{Metsaev:2009ym} and refs therein). Higher-spin conformal
(super)gravity theories have been further studied at the cubic
level by Fradkin and Linetsky in the frame-like formulation
\cite{Fradkin:1989xt
}
and a complete interacting theory has been proposed by Segal in
\cite{Segal:2002gd}. These theories may be thought of as the
higher-spin generalization of Weyl gravity. They are defined
around flat spacetime and contain higher derivatives, hence they
are non-unitary (but this is not an issue here since the gauge
fields are not dynamical). From the point of view of the
holographic correspondence, the effective action should be equal
to the on-shell action for the higher-spin gauge theory
around the AdS spacetime. Cubic vertices have been constructed by
Fradkin and Vasiliev \cite{Fradkin:1986qy
} (see
also the more recent works \cite{Vasiliev:2001wa
})
and full consistent equations of motion have been written by
Vasiliev \cite{Vasiliev:1990en
,Vasiliev:2003ev}.
These theories may be thought as higher-spin generalizations of
ordinary gravity. The existence of a conventional
variational principle for Vasiliev equations remains a major open
question. It is intriguing that all known tests of the
Klebanov-Polyakov conjecture circumvented successfully the lack of
a variational principle
\cite{Petkou:2003zz
,Giombi:2009wh,
Koch:2010cy,
Douglas:2010rc}.
Notice that the case of AdS${}_{3}$/CFT${}_{2}$  is special in this respect
\cite{Campoleoni:2010zq
}.
In brief, a detailed analysis of the regularized effective action
for free conformal scalars in a higher-spin background is
motivated both by the induced gravity program and by the AdS/CFT
correspondence. We shall come back to this issue in the
conclusion but let us  first summarize our results.

Consider a free complex massless scalar field, $\phi\,$, in flat
spacetime, described by the action
\be
    \mathcal{S}^{\sst [0]}[\phi] = -\int d^dx\,
    \phi^*\,\partial^2\phi\,
    = \la\,\phi\,|\,
    \hat{P}^2\,|\,\phi\,\ra\,.
\ee It has an infinite number of conserved currents given, for
instance, via the generating function
\be
    J(x,q)\,=\,\phi^*(x+q/2)\,\phi(x-q/2),
    \ee where we
introduced an auxiliary {vector} variable $q$ and the currents are
the Taylor coefficients of \be
 J(x,q)\,=\,\sum\limits_{s=0}^{\infty}\,
\frac 1{s!}\,J^{\sst (s)}_{\mu_1\dots\,\mu_s}(x)\,q^{\mu_1}\cdots
q^{\mu_s}\,. \label{gencurr} \ee These currents $J^{\sst
(s)}_{\mu_1\dots\,\mu_s}$ were first introduced in
\cite{Berends:1985xx}. They are conserved for massive scalar
fields as well, but the massless case is special because these
currents can be projected onto an infinite number of traceless
conserved currents, as explicitly performed below. Various
explicit sets of such conformal currents on Minkowski spacetime
were provided in
\cite{Anselmi:1999bb
}.

The scalar field can couple to external
higher-spin gauge fields, via the \emph{Noether coupling}:
\be
\mathcal{S}^{\sst (s)}_{\rm int}\big[J^{\sst (s)},h^{\sst (s)}\big]
\,=\,\, \int
d^dx \, \frac{(i\,\ell)^{s-2}}{s!}\,J^{\sst (s)}_{\mu_1\dots\,\mu_s}(x)\,
h^{{\sst (s)}\,\mu_1\dots\, \mu_s}(x)\,,
\label{Noether}
\ee
where $\ell$ is a coupling constant with the dimension of a
length and the powers of $i$ are such that their product with the currents (imaginary for $s$ odd) is real.
The external higher-spin field:
\be h^{\sst (s)}(x,u)\,=\,\frac1{s!}\,h^{\sst (s)}_{\mu_1\dots \mu_s}(x)\,u^{\mu_1}\dots
u^{\mu_s}\,,\ee is characterized by {a Fronsdal (like)
gauge symmetry \cite{Fronsdal:1978rb}}:
\be \delta^{\sst [0]} h^{\sst (s)}(x,u)=
(u\cdot\partial_x)\, \epsilon^{\sst (s-1)}(x,u)\,,\label{ga} \ee
where we introduced again an auxiliary variable $u$\,.
Under
this variation supplemented by a suitable linear
transformation $\delta\phi$ of the scalar field, the action
$\mathcal{S}^{\sst [0]}+\mathcal{S}^{\sst (s)}_{\rm int}$ 
is invariant up to terms of order $h$\,.

We showed in \cite{Bekaert:2009ud} that the sum $\mathcal{S}^{\sst
[0]}+\sum_{s=0}^{\infty} \mathcal{S}_{\rm int}^{\sst (s)}$ has an exact
symmetry group which reduces to  \eqref{ga} at lowest order. This
symmetry group is rendered manifest by first rewriting this sum
as \be \mathcal{S}^{\sst [0]}[\phi]+\sum_{s=0}^{\infty}\, \mathcal{S}_{\rm int}^{\sst(s)}[J^{\sst (s)},h^{\sst (s)}]\,=\, \la \,\phi\,|\, \hat{P}^2 -\ell^{-2}\, \hat{H}\,|\,\phi\,\ra\,, \ee
where the Hermitian operator $\hat H$ is given in terms of $h(x,u):=\sum_{s=0}^{\infty} h^{\sst (s)}(x,u)$ by \be \hat
H := \int \frac{d^{d}x\,
d^{d}p}{(2\pi)^{d}}\ h(x,\ell\,p)\ \delta(x-\hat X,p-\hat P)\,, \ee
with
\be
    \delta(\hat X,\hat P):=\int \frac{d^{d}y\,d^{d}k}{(2\pi)^{d}}\
    e^{i\,(k\cdot\hat X-y\cdot\hat P)}\,.
\ee
In other words, the generating function $h(x,\ell\,p)$ is the Weyl symbol
of the operator $\hat H$. The classical action is now manifestly
invariant under \be \delta\, |\,\phi\ra\,=\,\frac i2\,\hat
E\,|\,\phi\,\ra\,,\qquad \delta\, \hat H\,=\,\frac i2\,\big[\,\ell^{2}\,\hat
P^{2}-\hat H\,,\,\hat E\,\big]_{\sst -}\,, \ee where $\hat E$ is an
arbitrary infinitesimal Hermitian operator {and $[\,\,,\,]_{\sst
-}$ stands for the commutator}. In terms of the Weyl 
symbols $h(x,\ell\,p)$ and $\epsilon(x,\ell\,p)$ of, respectively, $\hat H$ and $\ell\,\hat E$, the
invariance reads \be \delta_\epsilon\,
h(x,u) = (u\cdot\partial_x)\,\epsilon(x,u)-
\frac i{2\,\ell}\,\Big[\epsilon(x,\ell\,p)\star h(x,\ell\,p)-
h(x,\ell\,p)\star\epsilon(x,\ell\,p)\Big]_{p=\frac{u}\ell}\,, \label{full} \ee where $\star$ is the
Moyal product (recalled in Appendix \ref{sec:Weyl}). This exact
symmetry is thus a deformation of the free gauge transformations to
which it reduces at the lowest order in $h$. It is also an extension
of the diffeomorphism and Maxwell gauge symmetries.

Actually, the massless classical action has a larger symmetry
group since the transformations \be \delta\,
|\,\phi\,\ra\,=\,\frac 12\,\hat A\,|\,\phi\,\ra\,,\qquad \delta\,
\hat H\,=\,\frac12\,\big[\,\ell^{2}\,\hat P^{2}-\hat H\,,\, \hat
A\,\big]_{\sst +}\,, \ee {where $\hat A$ is an infinitesimal
Hermitian operator and $[\,\,,\,]_{\sst +}$ stands for the
anticommutator,} leave the action invariant as well.  In terms of
symbols it reads \be \delta_\alpha\,
h(x,u)=(u^{2}-\tfrac{\ell^2}4\,\partial_{x}^{2})\,\alpha(x,u)-
\frac12\,\Big[ \alpha(x,\ell\,p)\star h(x,\ell\,p) + 
h(x,\ell\,p)\star \alpha(x,\ell\,p)\Big]_{p=\frac{u}\ell}\,, \label{Wlike} \ee
and so it represents a deformation
of a (generalized) Weyl
transformation \cite{Segal:2002gd}.\footnote{Such a deformation was analysed in AdS \cite{Manvelyan:2004mb}
 but only at lowest order and for external gauge fields of even spin.}
In particular, its linearization
is a deformation of the Weyl (like) gauge symmetries:
\be
    \delta^{\sst [0]}\, h^{\sst (s)}(x,u)\,=\,
    u^2\,\alpha^{\sst (s-2)}(x,u)\,,\label{gW}
\ee
mentioned in \cite{Fradkin:1985am}.
Consequently, the set of free gauge symmetries (Fronsdal and Weyl like) is deformed to the non-Abelian algebra of
differential operators acting on the scalar
field, where the Hermitian operators are associated with the deformation of the Fronsdal-like transformations \eqref{ga} while the anti-Hermitian operators are related to the deformation of the Weyl-like transformations \eqref{gW}. However, in general only the former classical symmetries are preserved at the quantum level while the
latter ones are anomalous (see \cite{Segal:2002gd} for an earlier discussion of these symmetries, though from a slightly different perspective). Via exponentiation, one may identify the symmetry group of a collection of free massless complex scalar fields in a higher-spin background as being: at classical level, the group of invertible (pseudo)differential operators and, at quantum level, the group of unitary operators (or symmetric operators if the scalar field is real).

The aim of this article is to calculate perturbatively the
quantum effective action of the scalar field in the higher-spin
background given by the Hermitian operator $\hat H$. {We introduce an ultraviolet (UV) cutoff
$\Lambda$ to make the effective action $\mathcal{W}_\Lambda[h]$ UV-finite.
A convenient regularization is the Schwinger proper time
regularization where $\mathcal{W}_\Lambda$ is given by the heat
kernel expansion: \be \mathcal{W}_\Lambda[h]=- \int_{1\over
\Lambda^2}^{\infty} {dt\over t}\ \mbox{Tr}\left[e^{-t(\hat
P^2-\hat H)}\right]\,. \ee We shall use perturbation theory in
the external fields (encoded in the generating function $h$) in
order to calculate $\mathcal{W}_\Lambda[h]$, and get the result in
the form: \be \mathcal{W}_\Lambda[h]=\mathcal{W}_{\rm fin}[h]\,
+\log\Lambda\  \mathcal{W}_{\rm log}[h]
+\sum_{n>-d/2}^{\infty}\Lambda^{2n+d}\ \mathcal{W}_n[h]\,+\,{\cal
O}(\Lambda^{-2})\,. \label{HKEW} \ee Each term in the above sum is
invariant under the full gauge transformations \eqref{full}. The
series is to be interpreted as an asymptotic series, the first
terms giving an approximate expression for  small fields. The
first term, $\mathcal{W}_{\rm fin}$, is a finite, cut-off
independent, non-local gauge-invariant expression. The second
term, $\mathcal{W}_{\rm log}[h]$, is the coefficient of the
logarithmically divergent piece which only appears when the
spacetime dimension $d$ is even, and it corresponds to the
higher-spin conformal gravity action proposed in
\cite{Segal:2002gd}. The remaining terms, $\mathcal{W}_n$, are the
coefficients of the UV divergent pieces, and they are sums of
local expressions which one can characterize as follows: each term
in $\mathcal{W}_n$ involves a certain {total number $p$ of
derivatives and a product of $m$ fields $h^{\sst (s_i)}$ such that
 \be
 2n = -p +\sum_{i=1}^m\,(s_i-2)\,.
 \ee
In particular, a term linear ($m=1$) in the field of spin $s_1$
appears in $\mathcal{W}_n$  for $s_1=2(n+1)\geqslant
0$.\footnote{Notice that imposing the condition
$\mathcal{W}_{-1}=0$ would determine, in a gauge invariant manner,
the external scalar $h^{\sst (0)}$ without derivative ($p=0$) in
terms of the other fields $h^{\sst (s>0)}$. This is the case for
free scalars fields in the presence of lower spins ($1\leqslant
s\leqslant 2$).} Such linear terms are invariant under the
linearized symmetries \eqref{ga} but not under the full ones
\eqref{full}. Consequently, inside such a $\mathcal{W}_n$ the
quadratic ($m=2$) terms in the fields with two derivatives ($p=2$)
are not invariant by themselves under the Fronsdal-like gauge
transformation \eqref{ga}. For these reasons, the quadratic
Fronsdal actions \cite{Fronsdal:1978rb} are \emph{not} recovered
for higher spins $s\geqslant 3$ inside the local UV divergent
terms (while the Maxwell and Pauli-Fierz can be recovered via a
suitable definition of the scalar $h^{\sst (0)}$).}

The finite term $\mathcal{W}_{\rm fin}$, the logarithmically
divergent term $\mathcal{W}_{\rm log}$, as well as the divergent
terms $\mathcal{W}_n$ for $n<-1$ start with quadratic terms in the
fields. This implies their invariance under the linearized gauge
transformations. We express these quadratic terms in a manifestly
invariant way by means of higher-spin curvatures. It turns out
that the deformed Weyl-like transformations \eqref{Wlike} do not
leave the whole effective action invariant but, when $d$ is odd,
the finite effective action is invariant and, when $d$ is even,
the logarithmic term is invariant. The quadratic actions of
\cite{Fradkin:1985am,Segal:2002gd} are reproduced as the quadratic
parts of $\mathcal{W}_{\rm fin}$ and $\mathcal{W}_{\rm log}$ in
the corresponding dimensions. As stressed by Segal, the
logarithmic term $\mathcal{W}_{\rm log}$ reproduces his non-linear
action \cite{Segal:2002gd} of conformal higher-spin gravity in
even dimensions. Interestingly, the finite term $\mathcal{W}_{\rm
fin}$ may be interpreted as providing a non-local action for
conformal higher-spin gravity in odd dimensions. 

The plan of this paper is as follows. In Section
\ref{sec:classical} we show that starting from the generating
function \eqref{gencurr} of currents, it is possible to construct
another one giving rise to on-shell traceless currents. The
advantage of working with such currents is that the linearized
Weyl-like gauge transformation is given exactly by the first term
in the right-hand-side of \eqref{gW} without corrections of order
$\ell^2$. The above-mentioned symmetries of the classical action
are examined in further details. 
In Section \ref{sec:quantum sym} 
the regularized one-loop effective action is introduced by
making use of the heat kernel
where its symmetries are discussed. More precisely, we show that the effective
action is not invariant under 
the transformations \eqref{Wlike} and we identify the generalized
Weyl anomalies. Section \ref{sec:comp} is devoted to a thorough analysis
of the regularized effective action through a double expansion:
first, the heat kernel is expanded as a power series of the
external fields, second the  trace of the heat kernel
is developed as a power series in the UV cutoff. 
The explicit expressions of all linear and quadratic terms 
are obtained in terms of special functions.
In Section \ref{sec:low}, the    examples of
the effective action for free scalar fields in the ordinary lower
spin fields background are revisited in the light of the previous
analysis. Section \ref{sec:conclu} is the conclusion where we
discuss potential domain of applications of our results. In
Appendix \ref{sec:not}, we introduce our notations by reviewing
several ingredients used in the text such as higher-spin
curvatures, Weyl-Wigner quantization and various special
functions. Some technical details of our computations have been
placed in Appendix \ref{sec:app comp}.

\section{Classical action and symmetries}
\label{sec:classical}

Consider $N$  free massless complex scalar fields $\phi^a\,$ with
$a=1,\dots,N$\,,  on the {conformal boundary of Euclidean  $AdS_{d+1}$
identified with the compactification of} $\mathbb{R}^d$\,. The
free action of the vector multiplet $\bm\phi=(\phi^a)$ is the
quadratic functional: \be
    \mathcal{S}[\bm\phi] =
    \int d^dx\ \partial_\mu\bm\phi^*(x)\bm\cdot\partial^\mu\bm\phi(x)\,,
\ee
where \be\bm\phi_1\bm\cdot\bm\phi_2:=\sum_{a=1}^N\,\phi_1^a\,\phi_2^a\,. \ee
This action is invariant under $U(N)$ (or $O(N)$ if the vector multiplet is real)
global transformations of the vector multiplet and
under the conformal transformations, for example under dilatation $\bm\phi$ transforms as
\be
    \bm\phi(\lambda\,x)=\lambda^{\Delta_{\bm\phi}}\,\bm\phi(x)\,,\qquad
    \Delta_{\bm\phi}=\frac{2-d}2\,.
\ee
The Euler-Lagrange equation is $\square\,\bm\phi(x)\approx 0$.
Equalities that are valid only on the mass shell will be denoted by a weak equality symbol $\approx$\,.

\subsection{Traceless conserved currents} \label{sec:traceless}

Since the theory is free, one can find infinitely many conserved
currents:  \be
    J^{\sst(s)}_{\mu_1\cdots\mu_s}(x)\,=\,\Big(\frac{1}{2}\Big)^s
    \sum_{n=0}^s\ (-1)^n\,\binom{s}{n}\
    \partial_{(\mu_1}\dots\partial_{\mu_n}\bm\phi(x)
    \bm\cdot \partial_{\mu_{n+1}}\dots\partial_{\mu_s)}\bm\phi^*(x)\,.
\label{currents}
\ee
From a simple generating function {(see Appendix \ref{genfct} for conventions)}:
\be
    J(x,q)= \bm\phi^*(x+\,q/2)\bm\cdot\bm\phi(x-\,q/2)\,.
    \label{genfct}
\ee
The equations of motion of $\bm\phi$ and $\bm\phi^*$ give the conservation condition,
$(\partial_x\cdot\partial_q)\,J(x,q)\approx0$\,, as well as another on-shell condition:
\be
    \left(\partial_q^2+\frac{1}{4}\,\partial_x^2\right) J(x,q)\approx 0\,.
    \label{2nd on-shell}
\ee The symmetric conserved current \eqref{currents} of rank $s$
is bilinear in the scalar field and contains exactly $s$
derivatives. The currents of odd rank are purely imaginary, thus
if the scalar field is real then they are absent. They were first
considered in \cite{Berends:1985xx} and obtained from the
generating function in \cite{Bekaert:2009ud}.

{Since the scalar fields are massless,} this free theory admits in addition the conformal symmetries.
Therefore, one may expect to get infinitely many \emph{traceless} conserved currents,
while the currents \eqref{currents} generated from \eqref{genfct} are not traceless,
even on-shell: $\partial_q^2\,J(x,q)\not\approx0$\,.
Nevertheless, thanks to the second on-shell condition \eqref{2nd on-shell}, one can construct
a generating function $\bar J(x,q)$ of conserved currents that are conserved and traceless on-shell:
\be
    \partial_q^2\,\bar J(x,q)\approx0\,,\qquad (\partial_x\cdot\partial_q)\,\bar J(x,q)\approx0\,,
    \label{traceless conservation}
\ee by acting with a differential operator
$\Pi_{d}(q,\partial_{x})$ on the generating function of
currents\,: $\bar J(x,q)=\Pi_{d}(q,\partial_{x})\,J(x,q)$\,. The
conservation of both $\bar J$ and $J$ requires that $\Pi_{d}$
commutes with $\partial_x\cdot\partial_q$ on-shell. If we
construct $\Pi_{d}$ as a power series in the operator
$P(q,\partial_{x}):=[(q\cdot\partial_x)^2-q^2\,\partial_x^2]/4$\,,
then the conservation condition is satisfied since
$\partial_x\cdot\partial_q\,P=P\,\partial_x\cdot\partial_q$\,. The
traceless condition,
\m{\partial_q^2\,\Pi_{d}(q,\partial_{x})\,J(x,q) \approx 0}\,,
remains to be solved and one needs to know how the trace acts on
powers of $P$\,: \ba
    && \partial_q^2\,P^n=\left[ P^n-4n\left(n-q\cdot\partial_{q}-\frac{d+1}{2}\right)\,
    P^{n-1}\right]\,\partial_q^2+ \nn
    && \quad +\,4n\left(n-q\cdot\partial_{q}-\frac{d+1}{2}\right)\,
    P^{n-1}\,\left(\partial_q^2+\frac{1}{4}\,\partial_x^2\right)
    +n\,P^{n-1}\,(q\cdot\partial_x)(\partial_x\cdot\partial_q)\,.
\ea By noting that the second line of the above equation vanishes
on-shell (when acting on $J$), one can construct an operator $\Pi_{d}$ as a power series in
$P$ with coefficients depending on the operator
$q\cdot\partial_{q}$ (counting the rank of the tensor fields it
acts on) such that all terms of $\partial_q^2\,\Pi_{d}$ cancel
each other on-shell. The operator $\Pi_{d}$ is determined by these
conditions (up to a constant factor): \be
    \Pi_d(q,\partial_{x}):= \sum_{n=0}^\infty\, \frac{1}{n!\,
    (-q\cdot\partial_{q}-\frac{d-5}{2})_{n}}\left(\frac{1}{4}\,P(q,\partial_{x})\right)^n\,,
    \label{Pi}
\ee where $(a)_{n}=\Gamma(a+n)/\Gamma(a)$ is the Pochhammer
symbol. More precisely, the operator $\Pi_d$ obeys  the identities:
\ba
    &(\partial_x\cdot\partial_q)\,\Pi_d=\Pi_{d+2}\,(\partial_x\cdot\partial_q)\,,\nn
    &\partial_q^2\,\Pi_{d}=\Pi_{d+4} \left(\partial_q^2+\frac{1}{4}\,\partial_x^2\right)
    -\frac{1}{2(d-1)+4\,q\cdot\partial_{q}}\,
    \Pi_{d+2}\,(q\cdot\partial_x)(\partial_x\cdot\partial_q)\,,
    \label{iden. Pi}
\ea so that it indeed fulfills the required properties. It will be
useful, for later use, to compute the inverse of the operator
$\Pi_{d}$, \textit{i.e.} the operator $\Pi^{-1}_{d}$ such that
$\Pi^{-1}_{d}\,\Pi_{d}=\Pi_{d}\,\Pi^{-1}_{d}=1$.   In Appendix
\ref{sec:inverse} we show that  \be
    \Pi^{-1}_{d}(q,\partial_{x})=\sum_{n=0}^{\infty} \left(\frac{1}{4}\,P(q,\partial_{x})\right)^{\!n}
    \frac1{n!\,(q\cdot\partial_{q}+\frac{d-1}2)_{n}}\,.
    \label{Pi'}
\ee

Finally, the explicit form of the generating function $\bar J(x,q)=\Pi_d(q,\partial_{x})\,J(x,q)$ of
traceless conserved currents is given by
\be
    \bar J(x,q)=\sum_{n=0}^\infty \frac{1}
    {n!\,(-q\cdot\partial_{q}-\frac{d-5}{2})_{n}}
    \left(\frac{(q\cdot\partial_x)^2-q^2\,\partial_x^2}{16}\right)^{\!n}\,
    \bm\phi^*(x+\,q/2)\bm\cdot\,\bm\phi(x-\,q/2)\,,
    \label{genfctcur}
\ee and by expanding in the variable $q^\mu$\,, one can find the
explicit form of the traceless currents. For instance, the spin $2$
traceless current reads \ba
    \bar J^{\sst(2)}_{\mu\nu}\e\frac{1}{4(d-1)}
    \Big[(d-2)\left(\partial_{\mu}\partial_{\nu}\bm\phi^{*}\bm\cdot\bm\phi
    +\bm\phi^{*}\bm\cdot\partial_{\mu}\partial_{\nu}\bm\phi\right) \nn
    && \hspace{50pt} -\,d\left(\partial_{\mu}\bm\phi^{*}\bm\cdot\partial_{\nu}\bm\phi
    +\partial_{\nu}\bm\phi^{*}\bm\cdot\partial_{\mu}\bm\phi\right)
	+\eta_{\mu\nu}\,\partial^{2}
    (\bm\phi^{*}\bm\cdot\bm\phi)\Big]\,.
\ea
Various explicit sets of conformal conserved currents on maximally symmetric spacetimes were also provided in \cite{Anselmi:1998bh,Vasiliev:1999ba,Anselmi:1999bb,Konstein:2000bi,Leonhardt:2004pp,Manvelyan:2004mb,Gelfond:2006be,Giombi:2009wh}.

\subsection{Scalar field action in a higher-spin background
 and its symmetries}
\label{sec:classical sym}

Now we couple each traceless currents $\bar J^{\sst (s)}$ to a
higher-spin gauge field $\bar h^{\sst (s)}$ and get the  boundary
action $\mathcal{W}[\bar h]$  as \be
    e^{-\mathcal{W}[\bar h]}=\int\mathcal{D}\bm\phi\
    \exp\!\left(-\mathcal{S}[\bm\phi]-\mathcal{S}_{\rm int}[\bar J,\bar h]\,\right)\,,
    \label{formal ea}
\ee
where $\mathcal{S}_{\rm int}$ is the collection of the Noether interactions
\eqref{Noether}:
\be
    \mathcal{S}_{\rm int}[\bar J,\bar h]:=\sum_{s=0}^{\infty}\,
        \mathcal{S}_{\rm int}^{\sst(s)}[{\bar J}^{\sst (s)},\,{\bar h}^{\sst (s)}]\,.
\ee
Since $\bar J(x,q)$ is conserved and traceless,
the action $\mathcal{S}[\bm\phi;\bar h]:=\mathcal{S}[\bm\phi]+
\mathcal{S}_{\rm int}[{\bar J},{\bar h}]$ is invariant,
when the equation of motion of $\bm\phi$ is implemented, 
under the gauge transformations:
\be
    \delta_{\bar \epsilon}\, \bar h(x,u)
    =(u\cdot\partial_x)\,{\bar \epsilon}(x,u)\,,\qquad
    \delta_{\bar \alpha}\, \bar h(x,u)=u^2\,{\bar \alpha}(x,u)\,,
    \label{linear transf}
\ee with arbitrary generating functions $\bar \epsilon$ and $\bar
\alpha$. In fact, these \emph{on-shell} symmetries can be deformed
to \emph{off-shell} symmetries when accompanied by a suitable
linear transformation of the scalar multiplet $\bm\phi$ together
with a higher order completion of \eqref{linear transf}. An easy
way to see this is to consider a new set of fields
$h(x,u):=\Pi_{d}(i\,\ell\,\partial_{u},\partial_{x})\,\bar h(x,u)$\,.
The relations between $h$ and $\bar h$ can be expressed
explicitly in terms of the different  components as\footnote{Note that \eqref{dressing} is similar to  the
\emph{dressing map} of \cite{Segal:2002gd} and  \eqref{undressing} is
another expression of the \emph{reconstruction map} of the same
article.} \ba
    h^{\sst (s)}(x,u) \e  \sum_{n=0}^{\infty}\,
    \frac{1}{n!\,(s+\frac{d-3}{2}+n)_{n}}
    \left(\ell^{2\,}\frac{(\partial_{u}\cdot\partial_x)^2
    -\partial_{u}^2\,\partial_x^2}{16}\right)^{n}\,
    \bar h^{\sst (s+2n)}(x,u)\,,
    \label{undressing} \\
    \bar h^{\sst (s)}(x,u) \e \sum_{n=0}^{\infty}
    \frac{(-1)^{n}}{n!\,(s+\frac{d-1}{2})_{n}}
    \left(\ell^{2}\,\frac{(\partial_{u}\cdot\partial_x)^2
    -\partial_{u}^2\,\partial_x^2}{16}\right)^{n}\,h^{\sst (s+2n)}(x,u)\,.
    \label{dressing}
\ea In terms of $h$, the coupling $\mathcal{S}_{\rm int}[\bar J,\bar h]$
becomes $\mathcal{S}_{\rm int}[J,h]$\,,
and the latter can be represented in a suggestive
way using the Weyl formalism (see Appendix \ref{sec:Weyl} for a
brief introduction). First, notice that  the Weyl symbol of
$\ket{\bm\phi}\!\bm\cdot\!\bra{\bm\phi}$ is given by \be
    \rho(x,p):=\int d^dq\, \bm\phi^{*}(x+q/2)\bm\cdot\bm\phi(x-q/2)\,e^{i\, q \cdot p}
    =\int d^dq\, J(x,\,q)\,e^{i\, q \cdot p}\,.
\ee Second, write the interaction with the external higher-spin gauge
fields as  \be
    {\ell}^{-2} \int \frac{d^{d}x\,d^{d} p}{(2\,\pi)^{d}}\,\rho(x,p)\, h(x,\ell\, p)\,.
\ee Finally, apply \eqref{star trace} to express the above
expression as \be
    {\ell}^{-2}\, {\rm Tr}\Big[\,\ket{\bm\phi}\!\bm\cdot\!\bra{\bm\phi}\,\hat H\,\Big]
    =\bra{\bm\phi} {\ell}^{-2}\, \hat H \ket{\bm\phi}\,,
\ee where  $\hat H$ is the Hermitian operator whose Weyl symbol is
$h(x,\ell\, p)$ and the summation on the vector
multiplet indices is understood. Now the classical action
$\mathcal{S}[\bm\phi; \bar h]$ reads \be
    \mathcal{S}[\bm\phi; \bar h]
    =\bra{\bm\phi} \hat P^2-\ell^{-2}\,\hat H\ket{\bm\phi}\,.
    \label{S class}
\ee
Defining $\hat G := \hat P^2-\ell^{-2}\,\hat H$\,,
the classical action \eqref{S class} is manifestly invariant under
\be
    \ket{\bm\phi}\ \to\ \hat{O}^{-1}\,\ket{\bm\phi}\,,\qquad
    \hat G\ \to\ \hat{O}^\dagger\,\hat G\,\hat O\,.
\ee
These symmetries are generated by two Hermitian operators $\hat A$ and $\hat E$\,:
\be
    \hat O=e^{-\frac{1}{2}(\hat A+i\,\hat E)}\,,
\ee
under which $\hat{H}$ transforms infinitesimally as
\be
    \delta_{\hat{E}}\,\hat{H} =
    \frac i2\,\big[\,\hat{G}\,, \hat{E}\,\big]_{\sst-}\ , \qquad
    \delta_{\hat{A}}\,\hat{H} = \frac12\,\big[\,\hat{G}\,,\hat{A}\,\big]_{\sst +}\,,
\ee
where $[\,\hat{A}\,,\hat{B}\,]_\pm:= \hat{A}\,\hat{B} \pm \hat{B}\,\hat{A}$ denotes the (anti)commutator, and  we hereafter set $\ell=1$\,.
The latter equations respectively read, in terms of Weyl symbols,
\ba
    \delta_\epsilon\,h(x,u) \e (u\cdot\partial_x)\,\epsilon(x,u)
    -\frac i2\,\big[\,h(x,u)\,\stackrel{\star}{,}\,\epsilon(x,u)\,\big]_{\sst-}\,,
    \label{nonlinear e transf} \\
    \delta_\alpha\,h(x,u) \e \left(u^2-\frac14\,\partial_x^2\right) \alpha(x,u)
    -\frac12\, \big[\,h(x,u)\,\stackrel{\star}{,}\,\alpha(x,u)\,\big]_{\sst+}\,,
    \label{nonlinear a transf}
\ea where $\epsilon(x,p)$ and $\alpha(x,p)$ are the respective
Weyl symbols of $\hat E$ and $\hat A$\,, and
\be
    \big[\,f(x,u)\,\stackrel{\star}{,}\,g(x,u)\,\big]_{\sst\pm}
    :=f(x,u)\star g(x,u)\pm g(x,u)\star f(x,u)
\ee denotes the Moyal (anti)commutator.
Notice that, in general, even if we start with a single $h^{\sst (s)}$, the
non-linear $\epsilon^{\sst (s-1)}$-transformations will generate
$h^{\sst (2s-2)}, h^{\sst (2s-4)}, \dots, h^{\sst (0)}$\,.
This higher-spin field generation also can be seen from the non-Abelian structure of gauge transformations:
\be
     \delta_{\epsilon_{3},\alpha_{3}}
     :=\big[\,\delta_{\epsilon_{1},\alpha_{1}}\,,\,\delta_{\epsilon_{2},\alpha_{2}}\,
     \big]_{\sst -}\,,
\ee
with
\ba
    \epsilon_{3}(x,u) \e
   \frac{i}{2}\, \big[\,\epsilon_{1}(x,u)\,\stackrel{\star}{,}\,\epsilon_{2}(x,u)\,\big]_{\sst-}
    -\frac{i}{2}\,\big[\,\alpha_{1}(x,u)\,\stackrel{\star}{,}\,\alpha_{2}(x,u)\,\big]_{\sst-}\,, \nn
    \alpha_{3}(x,u) \e
    \frac{i}{2}\,\big[\,\epsilon_{1}(x,u)\,\stackrel{\star}{,}\,\alpha_{2}(x,u)\,\big]_{\sst-}
    -\frac{i}{2}\,\big[\,\epsilon_{2}(x,u)\,\stackrel{\star}{,}\,\alpha_{1}(x,u)\,\big]_{\sst-}\,.
    \label{composition}
\ea
Even though we start with single spin gauge parameters
$\epsilon^{\sst (r)}$ and $\alpha^{\sst (q)}$,
they will automatically generate $\epsilon^{\sst (2p-1)}, \epsilon^{\sst (2p-3)}, \dots $
and $\alpha^{\sst (2p-1)}, \alpha^{\sst (2p-3)}, \dots $
where $p={\rm min}(r,q)$\,.
Thus there are only three consistent {spectra of external gauge fields that contain a finite
number of them}: $\{ h^{\sst (1)}, h^{\sst (0)} \}, \{ h^{\sst
(2)}, h^{\sst (0)} \}$ and $\{ h^{\sst (2)}, h^{\sst (1)}, h^{\sst
(0)}\}$, and two theories with an infinite number of fields: one
with all even spin fields $\{ h^{\sst (2 n)}\,|\,n\in \mathbb{N}
\}$ and the other with all fields $\{ h^{\sst (s)}\,|\, s\in
\mathbb{N} \}$. In fact, the three cases with a finite number of
fields can be viewed as complex scalar fields in the background of
electromagnetic and gravitational fields, when proper redefinition
of fields are carried out {(in particular the external scalar field $h^{\sst (0)}$ is determined in terms of the other fields).} Then the corresponding effective action
gives gauge invariants for each order of cut-off, among which one
can find in particular the Maxwell action and the Einstein-Hilbert
action. We will come back in Section \ref{sec:low} to this point
and  review some aspects of the ordinary lower spin cases.

 The gauge parameters $(\epsilon, \alpha)$ are given in terms of 
 $(\bar \epsilon,\bar\alpha)$ using \eqref{linear transf},  
 the $h$-independent part of 
 (\ref{nonlinear e transf}$-$\ref{nonlinear a transf})  
  and \eqref{iden. Pi}   as \ba
    \epsilon(x,u) \e  \Pi_{d+2}(i\,\partial_{u},\partial_{x})\ \bar \epsilon(x,u)
    +(\partial_x\cdot\partial_u)\ \Pi_{d+2}(i\,\partial_{u},\partial_{x})\
    \frac{1}{2(d-1)+4\,u\cdot\partial_{u}}\,\bar \alpha(x,u)\,,\nn
    \alpha(x,u)\e\Pi_{d+4}(i\,\partial_{u},\partial_{x})\ \bar \alpha(x,u)\,.
    \label{gauge par iden}
\ea Thus, \eqref{nonlinear e transf}, \eqref{nonlinear a transf}
and \eqref{gauge par iden} define the completion of the
free gauge transformation \eqref{linear transf} of $\bar h$ with
gauge parameters $\bar \epsilon$ and $\bar \alpha$\,.
Notice also that the gauge parameters which do not
generate any free gauge transformation:
\be
    (u\cdot\partial_{x})\,\bar\epsilon(x,u)+u^{2}\,\bar\alpha(x,u)=0 \,,
    \label{conf killing}
\ee
define the conformal Killing tensors
(see \textit{e.g.} \cite{Eastwood:2002su} and refs therein).
One can check that for $d>2$ the lower spin solutions
($\bar\epsilon^{\sst (1)}$ and $\bar\alpha^{\sst (0)}$)
are related to the generators of the conformal algebra $\mathfrak{so}(d,2)$
(for $d=2$, one gets two copies of the Witt algebra, \textit{i.e.} the Virasoro algebra with vanishing central charge).
For the higher-spin case, the space of non-trivial solutions 
of \eqref{conf killing} for $d>2$ corresponds to
the enveloping algebra of $\mathfrak{so}(d,2)$\,, 
that is the higher-spin algebra identified by Eastwood and Vasiliev \cite{Eastwood:2002su,Vasiliev:2003ev} (for $d=2$, one gets two copies of the ${\cal W}_\infty$ algebra
with zero central charge \cite{Bekaert:2007mi}).
The relation between the  Eastwood-Vasiliev algebra and the gauge
transformations \eqref{nonlinear e transf}-\eqref{nonlinear a
transf} should be the higher-spin generalization of the relation
between the conformal algebra and the diffeomorphisms and Weyl
transformations of the metric. 

The $\bar\epsilon$-transformations are different from the
Fronsdal's gauge symmetries in that the gauge parameters are not
constrained to be traceless. The $\bar \alpha$-transformation is a
higher-spin generalization of the Weyl transformation. In the next
section, we will investigate whether these symmetries extend their
validity to the quantum level.

\section{Regularized  effective action and its symmetries} \label{sec:quantum sym}

The path integral representation \eqref{formal ea} of the
effective action can be written formally as \be
    \mathcal{W}[\bar h]=N\,{\rm Tr}\!\left[ \log\hat G\right]\,,
    \qquad \hat G:= \hat P^{2}-\hat H\,,
    \label{formal EA}
\ee and the trace, ${\rm Tr[\cdot]}=\int \bra{p}\cdot\ket{p}
d^{d}p/(2\pi)^{d}$\,, leads to  a divergent integral. Thus, one
should properly regularize the effective action \eqref{formal EA}.
It is crucial that the regularization should preserve the
classical gauge symmetries \eqref{nonlinear e transf}.
In the following, we use  the gauge-invariant Schwinger proper time
regularization. It consists in replacing in the effective action
$\log z$ by a regularized integral
$-\int_{1/\Lambda^{2}}^{\infty} (dt/t)\,e^{-t\,z}$\,: \be
    \mathcal{W}_{\rm reg}[\bar h|\Lambda] := -
    N\,{\rm Tr}\left[\int^\infty_{\frac1{\Lambda^2}}
    \frac{dt}{t}\,e^{-t\,\hat G}\right],\label{Wreg}
\ee where $\Lambda$ is a regularization parameter
of mass dimension. For the study of gauge symmetries of
 $\mathcal{W}_{\rm reg}[\bar h|\Lambda]$,
it will be useful to define the ($\hat A$-inserted) trace of the
heat kernel (or simply heat trace): \be
    \mathcal{K}[g,\alpha|t]:= {\rm Tr}\Big[e^{-t\,\hat G}\,\hat A\Big]
    \qquad [\mathrm{Re}(t)>0]\,,
    \label{heat trace with A}
\ee where the generating functions $g(x,u)$ and $\alpha(x,u)$ are
again the Weyl symbols of $\hat G$ and $\hat A$, in particular
$g(x,p)=p^{2}-h(x,p)$. Then, the regularized effective action is
given as an integral of the heat trace: \be
    \mathcal{W}_{\rm reg}[\bar h|\Lambda]:=- N\int^\infty_{\frac1{\Lambda^2}}
    \frac{dt}{t}\,\mathcal{K}[g|t]\,,
    \label{reg. schwinger}
\ee
where $\mathcal{K}[g|t]:= \mathcal{K}[g,1|t]={\rm Tr}\Big[e^{-t\,\hat G}\Big]$.
In fact the $\hat A$-inserted heat trace can be obtained from the
variation of the usual heat trace under $\delta\, \hat G=\hat A$
since \be \delta\,{\rm Tr}\!\left[ e^{-t\,\hat G}\right]=
   - {\rm Tr}\left[ \int_{0}^{t} d\tau\,e^{-\tau\,\hat G}\,\hat A\,
    e^{-(t-\tau)\,\hat G}\right]
    = -t\, \mathcal{K}[g,\alpha|t]\,.
\ee The functional derivative with respect to $h$ is now
conveniently defined as \be 
\delta\, \mathcal{K}[g|t]=\ppd\alpha{ \tfrac{\delta}{\delta
h}}\,\mathcal{K}[g|t]\,, \ee
 where the double bracket $\ppd{f}{g}$
means the integral of the contraction between two generating
functions $f$ and $g$ (see Appendix \ref{genfct} for the precise
definition). Finally the relation between
$\mathcal{K}[g,\alpha|t]$ and $\mathcal{K}[g|t]$ can be written as
\be \mathcal{K}[g,\alpha|t]=-{1\over t}\, \ppd\alpha{
\tfrac{\delta}{\delta h}}\,\mathcal{K}[g|t]\,. \ee

The heat trace $\mathcal{K}[g,\alpha|t]$
can be expanded asymptotically (see Section \ref{sec:HKE}) as
\be
    \mathcal{K}[g,\alpha|t] =
    t^{-\frac{d}{2}}\sum_{n=-\infty}^{+\infty} t^n\,a_n[g,\alpha]\,,
    \label{K}
\ee
where the $a_n[g,\alpha]$ are ($\hat A$-inserted) heat kernel expansion (HKE) coefficients,
and  we will see in the following
 that the dimensional dependency $t^{-d/2}$ of the above expansion
 is essential in the analysis of the gauge symmetries.
 We can again get the $\hat A$-inserted HKE coefficients
 from the usual HKE coefficients $a_n[g]:=a_n[g,1]$ as
 \be
    a_{n}[g,\alpha]=- \ppd\alpha{\tfrac{\delta}{\delta h}}\,a_{n+1}[g]\,.
    \label{diff a}
\ee
Notice that the above equation relates two different kind HKE coefficients, one
with $\hat A$ insertion and the other without, but more importantly it relates
also different $n$'s.
In fact if we consider the HKE coefficient densities $b_{n}[g](x)$\,:
\be
    a_{n}[g]=:\int d^{d}x\ b_{n}[g](x)\,,
\ee
then by replacing $\alpha$  by a delta distribution in \eqref{diff a},
we get
\be
    b_{n}[g](x) = - \frac{\delta\,a_{n+1}[g]}{\delta\, h^{\sst (0)}(x)}\,.
    \label{rel heat}
\ee
As far as we know, this relation between two neighboring HKE coefficients was
not known before, and we can understand why it was difficult to notice in
the studies of the lower spin background fields.
That is because in those cases
$h^{\sst (0)}$ is usually given by a nonlinear function of other external fields.
Even in that case, we may keep $h^{\sst (0)}$ as independent, and at the end of computation replace it in terms of other fields.

 The regularized effective action itself can be expressed using the HKE coefficients
 as in the expansion \eqref{HKEW}:
\be
    \mathcal{W}_{\rm reg}[\bar h|\Lambda]
    =\mathcal{W}_{\rm fin}[\bar h|\mu]-
    2\,N\,\ln(\Lambda/\mu)\,a_{\frac{d}{2}}[g]-
    N\,\Lambda^{d} \hspace{-12pt}
    \sum_{n\neq \frac{d}{2},\, n=-\infty}^{+\infty} \!\!\!
    \Lambda^{-2n}\,\frac1{\frac{d}2-n}\,a_n[g]\,,
    \label{EA expansion}
\ee where $\mu$ is a constant of mass dimension, and the
coefficient $a_{d/2}[g]$ is non-vanishing only for even $d$\,. 
The finite part of the effective action $\mathcal{W}_{\rm fin}[\bar h|\mu]$
 is not given directly by a HKE coefficient but
requires the evaluation of
 the integral \eqref{reg. schwinger} and receives contributions from
 all the coefficients $a_{n}[g]$
with $n\geqslant d/2$\,.

Now we can examine the gauge symmetries of the regularized
effective action via $\mathcal{K}[g|t]\,$. The
$\epsilon$-transformation \eqref{nonlinear e transf} corresponds
to the adjoint action $\delta_{\hat{E}}\,\hat{G} =-\frac
i2\,[\,\hat{G}\,{,}\,\hat{E}\,]_{\sst-}$ and manifestly leaves the
trace $\mathcal{K}[g|t]$ invariant for all $t$\,, so  the HKE
coefficients $a_n[g]$ as well as the finite part of the effective
action $\mathcal{W}_{\rm fin}[\bar h|\mu]$ are invariant under
this gauge transformation as they should. On the other hand, the
variation of $\mathcal{K}[g|t]$ under the $\alpha$-transformation
\eqref{nonlinear a transf} does not vanish but gives \be
    \delta_\alpha\,\mathcal{K}[g|t]={\rm Tr}\Big[\Big(-t\,e^{-t\,\hat G}\Big)
    \Big(-\frac12 \,\big[\,\hat{G}\,{,}\,\hat{A}\,\big]_{\sst+} \Big) \Big]
    =-t\,\frac{d\, }{dt}\,\mathcal{K}[g,\alpha|t]\,,
    \label{A-transf K}
\ee
or  equivalently
\be
    \delta_\alpha\,a_n[g]=\left(\frac{d}2-n\right) a_n[g,\alpha]\,.
\ee
One can see that there is no $\alpha$-invariant HKE coefficient for odd $d$\,,
while for even $d$ the only invariant is $a_{d/2}[g]$ which gives a logarithmically divergent term
in $\mathcal{W}_{\rm reg}[\bar h|\Lambda]$\,.
Plugging \eqref{A-transf K} directly to \eqref{reg. schwinger}, we get
the gauge variation of the regularized effective action:
\be
    \delta_\alpha\,\mathcal{W}_{\rm reg}[\bar h|\Lambda]
    =-N\,\mathcal{K}[g,\alpha|\Lambda^{-2}]\,,
    \label{compare}
\ee which can be expanded using the HKE coefficients. Comparing
the expansion of this gauge variation \eqref{compare} with the gauge variation of
the expansion \eqref{EA expansion}, we get finally
\be
    \delta_\alpha\,\mathcal{W}_{\rm fin}[\bar h|\mu]
    =: \ppd{\alpha}{\mathscr{A}[\bar h]}
    =-N\,a_{\frac{d}{2}}[g,\alpha]\,.
    \label{gauge var. fin EA}
\ee Thus, for odd $d$ the finite part of the effective action is
invariant under the $\alpha$-transformation, but for even $d$ the
symmetry is anomalous. When the gauge parameter $\alpha$ is a
scalar, that is $\alpha=\alpha^{\sst (0)}(x)$\, with no $u$
dependence, the variation of the finite part of the effective action, that is the 
Weyl anomaly, is given by the Weyl invariant
HKE coefficient density:
\be
	\mathscr{A}^{\sst(0)}[\bar h](x)=
	N\,\frac{\delta\,a_{\frac d2+1}[g]}{\delta\,h^{\sst(0)}(x)}=
	-N\,b_{\frac d2}[g](x)\,,
	\label{Weyl anomaly}
\ee
which also corresponds to
the logarithmically divergent part of the effective action. If
the gauge parameter is generic, say $\alpha=\alpha^{\sst (r)}$
then it is not given by one of HKE coefficient densities but by
the generalized Weyl anomaly: 
\be
    \mathscr{A}^{\sst(r)}[\bar h](x,u)=
    N\,\left(\frac{\delta\ \ }{\delta h^{\sst (r)}}\,a_{\frac d2+1}[g] \right)\!(x,u)\,.
    \label{HS Weyl anomaly}
\ee

\section{Perturbative calculation of the effective action}
\label{sec:comp}

Up to now, we have considered the free scalar theory on the
boundary as the conjectured dual of an interacting higher-spin gauge theory
in AdS, and analyzed the gauge symmetries of the effective action
$\mathcal{W}[\bar h]$ from the scalar theory with a proper
regularization. The finite part of the effective action $\mathcal{W}_{\rm
fin}[\bar h|\mu]$, according to the AdS/CFT correspondence should
correspond in the semiclassical regime to the on-shell evaluation
of the action of the higher-spin gauge theory in AdS.

In this section, we compute the regularized effective action
$\mathcal{W}_{\rm reg}[\bar h]$ via the calculation of the trace
of the heat kernel $\mathcal{K}[g|t]$. In Section \ref{sec:comp.
heat} we will reduce the expression for the trace of the heat kernel
down to a Gaussian integral, and
in Section \ref{sec:HKE}, by evaluating this integral,
we obtain the HKE coefficients as well-defined multiple integrals.
In Section \ref{sec:quad
heat} we obtain explicit formulae 
for the linear and quadratic parts of the HKE coefficients.
In Section
\ref{sec:finite}, by integrating the heat trace, we get the
finite part of the effective action, up to quadratic order in fields, composed 
of a non-local part and, if the
dimension is even, also a local part. For $d>2$, the non-local part can be
rewritten as the generating functional of the connected correlation functions 
while the local part is not invariant
under the generalized Weyl transformation.  Finally in Section
\ref{sec:correl} we present an alternative way to compute all the
correlation functions .

\subsection{Trace of the heat kernel}
\label{sec:comp. heat}

The trace of the heat kernel $\mathcal{K}[g|t]$ can be computed as
a perturbation series in $h$ by expanding the heat kernel as \be
    e^{-t\,\hat P^{2}+t\,\hat H}=
    e^{-t\,\hat P^{2}}\,
    \sum_{n=0}^{\infty}\,
    \int_{0}^{t} d\tau_{1} \int_{0}^{\tau_{1}} d\tau_{2}
    \cdots \int_{0}^{\tau_{n-1}} d\tau_{n}\
    \hat H(\tau_{1})\,\cdots\,\hat H(\tau_{n})\,,
\ee
with $\hat H(t):= e^{t\,\hat P^{2}}\,\hat H\,e^{-t\,\hat P^{2}}$.
Replacing the trace by multiple integrals over $p$ in the cyclicly symmetric way
and replacing the matrix elements of $\hat H$ by integrals over $x$\,:
\be
    \bra p \hat H \ket q=\int d^dx\ h(x,\partial_u)\,
    e^{\frac12\,u\cdot(p+q)-i\,(p-q)\cdot x}\,\big|_{u=0}\,,
    \label{op H to fun H}
\ee
we can express the heat trace as
\be
    \mathcal{K}[g|t]=
    \sum_{n=0}^\infty\,\ppd{K^{\sst (n)}(t)}{h^{\otimes n}}\,,
    \label{K with kernel}
\ee
where $K^{\sst (n)}(t)$ is given by
\ba
    && K^{\sst (n)}(x_1,u_1;\cdots;x_n,u_n|t)= \nn
    &&=\left[\prod_{m=1}^{n}\,\int \frac{d^dp_m}{(2\pi)^d}\
    e^{i\,p_m\cdot[x_{m-1}-x_m-i\,(u_{m-1}+u_m)/2]}\right]
    \tilde K^{\sst (n)}(p_1,\cdots,p_n|t)\,,
    \label{K to tilde K}
\ea
with $x_{0}= x_{n}$\,, $u_0= u_n$\,, and $\tilde K^{\sst (0)}(t)= K^{\sst (0)}(t)=1$
and for $n\geqslant1$
\ba
    && \tilde K^{\sst (n)}(p_1,\cdots,p_n|t) =
    \int_0^{t} d\tau_1 \int_0^{\tau_1}d\tau_2 \cdots
    \int_0^{\tau_{n-1}}d\tau_n\ \frac{1}{n}\,\sum_{\ell=1}^n\,\times \\
    && \ \times
     \exp\Big[(\tau_1-t)\,p_\ell^2+\cdots
     +(\tau_{n-\ell+1}-\tau_{n-\ell})\,p_n^2+ \nn
     &&\qquad\qquad +\ (\tau_{n-\ell+2}-\tau_{n-\ell+1})\,p_1^2+\cdots +
     (\tau_n-\tau_{n-1})\,p_{\ell-1}^2 -\tau_n \,p_{\ell}^2)\Big] \,.
     \nonumber
     \label{tilde K}
\ea The computation of the heat trace  is reduced to that of
$\tilde K^{\sst (n)}(t)$ where, after changing variables to
$\sigma_m=\tau_{m-1}-\tau_m$ with $\tau_0=t$\,, the time-ordered
integral becomes \ba
    && \int_0^{t} d\tau_1 \int_0^{\tau_1}d\tau_2 \cdots
     \int_0^{\tau_{n-1}}d\tau_n
    = \int_0^\infty d\sigma_1\cdots \int_0^\infty
     d\sigma_n\ \Theta(t-\sigma_1-\cdots-\sigma_n) \nn
    && \qquad =\,\int_0^\infty d\sigma_0 \int_0^\infty
    d\sigma_1\cdots \int_0^\infty d\sigma_n\
    \delta(\sigma_0+\sigma_1+\cdots+\sigma_n-t)\nn
    && \qquad =\,\int_{-\infty}^{\infty}\frac{d\omega}{2\pi}\,e^{i\,\omega\,t}\,
        \int_0^\infty d\sigma_0\,e^{-i\,\omega\,\sigma_0} \cdots
        \int_0^\infty d\sigma_n e^{-i\,\omega\,\sigma_n}\,,
    \label{time order}
\ea then we
 calculate the integrals over $\sigma$\,: $\int_0^\infty
d\sigma\,e^{-\sigma(p^2+i\omega)}=\frac1{p^2+i\,\omega}$\,, we get
\ba
    \tilde K^{\sst (n)}(p_1,\cdots,p_n|t) \e
    \int_{-\infty}^{\infty}\frac{d\omega}{2\pi}\,e^{i\,\omega\,t}\,\frac{1}{n}
    \left(\sum_{\ell=1}^n \frac{1}{p_\ell^2+i\,\omega}\right)
    \frac{1}{(p_1^2+i\,\omega)\cdots (p_n^2+i\,\omega)} \nn
    \e \frac{t}{n}\int_{-\infty}^{\infty}\frac{d\omega}{2\pi}\,e^{i\,\omega\,t}\,
    \frac{1}{(p_1^2+i\,\omega)\cdots (p_n^2+i\,\omega)}\nn
    \e \frac{t}{n}\,\sum_{m=1}^{n}\,
     \frac{e^{-t\,p_m^2}}{\prod_{\ell=1,\ell\neq m}^n
     (p_\ell^2-p_m^2)}\,,
    \label{resultat tilde K}
\ea where we used an integration by part and the residue theorem for the evaluation
of the integral over $\omega$\,.
As a function of a $p_{\ell}^{2}$, \eqref{resultat tilde K}
is a sum of $n-1$ rational functions with a single pole and an
exponential divided by a polynomial with $n-1$ zeros. Because of
these polynomials, it seems to have many poles arising when other
momenta approach to $p_{\ell}$, but in fact there are no such poles
since they are all compensated by the poles of the rational
functions. 
Therefore, \eqref{resultat
tilde K} is a sum of Gaussian functions  of $p_{\ell}$ multiplied
by a series with only non-negative powers of $p_{\ell}$.

By using \eqref{K to tilde K} and \eqref{resultat tilde K}, one can eventually
compute $K^{\sst (n)}(t)$, that is the non-local representation of the heat trace.
In the following we will concentrate on the local representation.

\paragraph{Local functional representation}

In \eqref{K with kernel}, the heat trace is given with $K^{\sst
(n)}(t)$'s which are functions of differences between position
variables, $x_{m-1}-x_m$\,. If we integrate out all these
variables, the heat trace can be represented as a local
functional. To do so, let us first focus on the $n$-th order term:
\ba
    &&\ppd{K^{\sst (n)}(t)}{h^{\otimes n}}=
    \int d^{d}x_{1}\cdots d^{d}x_{n}\,\frac{d^{d}p_{1}}{(2\pi)^{d}}\cdots
    \frac{d^{d}p_{n}}{(2\pi)^{d}}\, \tilde K^{\sst (n)}(p_{1},\cdots,p_{n}|t)\times \nn
    && \ \ \times
    \left< \!\exp\!\left(i\,\sum_{\ell=1}^{n}\,p_{\ell}\!\cdot\!\left[x_{\ell-1}-x_{\ell}-
    \frac i2\,(u_{\ell-1}+u_{\ell})\right]\!\right)\!\bigg|\,
    h(x_{1},u_{1})\,\cdots\,h(x_{n},u_{n})\right>.
\ea
By defining new variables
\be
    x=\frac 1n\left( x_{1}+\cdots +x_{n}\right),
    \qquad
    y_{\ell}=x_{\ell}-x_{\ell-1}\,,
\ee
we can Taylor expand $h$ around the center position $x$, with $y_{\ell+n} = y_{\ell}$, as
\be
    h\big(x_{\ell}(x,\{y_{k}\}),u_{\ell}\big)
    =\exp\left[\frac1n\,\sum_{m=0}^{n}\left(m-\frac{n+1}2\right) y_{m+\ell}
    \cdot \partial_{x_{\ell}} \right] h(x_{\ell},u_{\ell})\,\Big|_{x_{\ell}=x}\,,
\ee
and replace the integration measure as
\be
    d^{d}x_{1}\cdots d^{d}x_{n} =
    d^{d}x\,d^{d}y_{1}\cdots d^{d}y_{n}\,\int \frac{d^{d}q}{(2\pi)^{d}}\,
    e^{i\,q\cdot(y_{1}+\,\cdots+\,y_{n})}\,.
\ee Then we can perform the integral over $y_{\ell}$ and get a delta
function which removes the integral over $p_{\ell}$ with
$p_{\ell}(q)=q-i\,\partial_{y_{\ell}}$ where \be
    \partial_{y_{\ell}}=\frac1n\,\sum_{m=1}^n\left(m-\frac{n+1}{2}\right)\,
    \partial_{x_{\ell-m}}\qquad [x_{\ell+n}=x_{\ell}]\,.
    \label{p in q}
\ee
Finally we get a local functional representation:
\be
    \ppd{K^{\sst (n)}(t)}{h^{\otimes n}}
    = \int d^d x\ V^{\sst (n)}(\partial_{x_1},\partial_{u_1};\cdots;
    \partial_{x_n},\partial_{u_n}|t)\,h(x_1,u_1)\cdots h(x_n,u_n)\,
    \Big|_{\overset{x_\ell=x}{\sst u_\ell=0}}\,,
    \label{V}
\ee
with
\ba
    && V^{\sst (n)}(\partial_{x_1},\partial_{u_1};\cdots;
    \partial_{x_n},\partial_{u_n}|t) = \int \frac{d^dq}{(2\pi)^d}\
    \tilde K^{\sst (n)}(p_1(q),\cdots,p_n(q)|t)\,\times
     \nn
    && \hspace{100pt} \times\,
    \exp\Big(\frac{p_n(q)+p_1(q)}2\cdot\partial_{u_1}
    + \cdots +
    \frac{p_{n-1}(q)+p_n(q)}2\cdot\partial_{u_n}\Big)\,.\qquad
    \label{V n}
\ea
The $n=0, 1$ cases can be immediately computed as
\be
    V^{\sst (0)}(t)=(4\pi\,t)^{-\frac{d}{2}}\,,
    \qquad
    V^{\sst (1)}(\partial_{x},\partial_{u}|t)
    =(4\pi\,t)^{-\frac{d}{2}}\,t\,
    e^{\frac{1}{4t}\,\partial_{u}^2}\,.
    \label{linear}
\ee From the discussion made below \eqref{resultat tilde K}, one
can see that $V^{\sst (n\geqslant 2)}$ has a form of a Gaussian integral
multiplied by a $q$-series with only non-negative powers, which
can be evaluated order by order in $q$. Notice also that the order
of $q$ in the $q$-series is equal to the sum of the number of
derivatives and the number of total spin. Therefore if we want to
compute the heat trace up to a fixed number of  derivatives or
total spin, then it is sufficient to consider the $q$-series to
that order, and to evaluate the Gaussian integrals. In the next
section, we will rather consider the expansion in $t$ than in $q$,
which gives the HKE coefficients as integrals.

\subsection{Heat kernel expansion}
\label{sec:HKE}

In the preceding section, we obtained the trace of heat kernel as
a Gaussian integral multiplied by a function which can be expanded
as a $q$-series. In the present section, we will evaluate the
Gaussian integral, without expanding in $q$\,, 
by noticing that $\tilde K^{\sst (n)}(t)$ \eqref{resultat tilde K} 
can be written as an $n$-ple integral: \be
    \tilde K^{\sst (n)}(p_1,\cdots,p_n|t)= \frac{t^{n}}{n}\int_{0}^{1}
    d\rho_{1}\cdots d\rho_{n}\,
    \delta(\rho_{1}+\cdots +\rho_{n}-1)\,
    e^{-t \left( \rho_{1}\,p_{1}^{2} +\cdots +\rho_{n}\,p_{n}^{2}\right)} .
\ee
By plugging this expression into \eqref{V n} and by changing the order of integrals,
one can evaluate the Gaussian integral and get for $n\geqslant2$
\ba
 \label{VV n}
    && V^{\sst (n)}(\partial_{x_1},\partial_{u_1};\cdots;
    \partial_{x_n},\partial_{u_n}|t) =
    \frac{t^{n}}n\,(4\pi\,t)^{-\frac d2}
    \int_{0}^{1}
    d\rho_{1}\cdots d\rho_{n}\,
    \delta(\rho_{1}+\cdots +\rho_{n}-1)\,\times \nn
    && \ \times\,
    \exp\Bigg[t \left(\, \left( \rho_{1}\,\partial_{y_{1}}^{2}+\cdots+
    \rho_{n}\,\partial_{y_{n}}^{2}\right)
    -\left( \rho_{1}\,\partial_{y_{1}}+\cdots+
    \rho_{n}\,\partial_{y_{n}}\right)^{2} \right)
    +\frac1{4\,t}\,\partial_{u_{1\sim n}}^{2} + \\
    && \qquad\qquad +\,
    i \left( \rho_{1}\,\partial_{y_{1}}+\cdots+
    \rho_{n}\,\partial_{y_{n}}\right)\cdot \partial_{u_{1\sim n}}
    -i \left(\partial_{y_{1}} \cdot\frac{\partial_{u_{1}}+\partial_{u_{2}}}2
    +\cdots+\partial_{y_{n}} \cdot\frac{\partial_{u_{n}}+\partial_{u_{1}}}2\right)
    \Bigg]\,, \nonumber
\ea where we used the notation \eqref{p in q} and
\m{\partial_{u_{1\sim
n}}:=\partial_{u_{1}}+\cdots+\partial_{u_{n}}}. The second line of
\eqref{VV n} has the form of the generating function \eqref{bessel
gen} of modified Bessel function $I_{m}(z)$, so by expanding in
$t$, we get $V^{\sst (n)}$ for $n\geqslant 2$ as an infinite series: \ba
    && V^{\sst (n)}(\partial_{x_1},\partial_{u_1};\cdots;
    \partial_{x_n},\partial_{u_n}|t) = \nn
    && \quad =
    \frac{t^{n}}n\,(4\pi\,t)^{-\frac d 2}
    \sum_{m=-\infty}^{\infty} t^{m}\,
    V^{\sst (n)}_{m}\!\left(\partial_{y_{1}},\frac{\partial_{u_{1}}
    +\partial_{u_{2}}}2;\cdots
    ;\partial_{y_{n}},\frac{\partial_{u_{n}}+\partial_{u_{1}}}2\right),
    \label{Vn expansion}
\ea
with
\ba
	       \label{Q n m}
    && V^{\sst (n)}_{m}(\partial_{y_{1}},\partial_{u_{1}};\cdots;
    \partial_{y_{n}},\partial_{u_{n}})=e^{-i (\partial_{y_{1}}\!\cdot \partial_{u_{1}}
    +\,\cdots\,+ \,\partial_{y_{n}}\!\cdot \partial_{u_{n}})} \times\nn
    &&\quad \times
    \int_{0}^{1} d\rho_{1}\cdots d\rho_{n}\,\delta(\rho_{1}+\cdots +\rho_{n}-1)\
     e^{i \left( \rho_{1}\,\partial_{y_{1}}+\,\cdots\,+
    \rho_{n}\,\partial_{y_{n}}\right)\cdot \partial_{u_{1\sim n}}}\,\times\\
    &&\quad \times
    \Big[2\,f_{n}(\rho_{1},\cdots,\rho_{n};\partial_{y_{1}},\cdots,\partial_{y_{n}})
    \Big]^{m}\,
    U_{m}\Big(-4\,f_{n}(\rho_{1},\cdots,\rho_{n};\partial_{y_{1}},\cdots,\partial_{y_{n}})\,
   \partial_{u_{1\sim n}}^{2}\Big),\nonumber
\ea
where we used the definition:
\be
    U_{\nu}(z) := (\sqrt{z}/2)^{-\nu} J_{\nu}(\sqrt{z}/2)
    =\sum_{n=0}^{\infty}
    \,\frac{1}{n!\,\Gamma(\nu+n+1)\,2^{\nu}} \left(-\frac{z}{16}\right)^{n}
    \,,
    \label{U bessel}
\ee
and
\be
    f_{n}(\rho_{1},\cdots,\rho_{n};\partial_{y_{1}},\cdots,\partial_{y_{n}}):=
    \rho_{1}\,\partial_{y_{1}}^{2}+\cdots+\rho_{n}\,\partial_{y_{n}}^{2}
    -\left(\rho_{1}\,\partial_{y_{1}}+\cdots+\rho_{n}\,\partial_{y_{n}}\right)^{2}\,.
    \label{ff}
\ee
Since $z^{m}\,U_{m}(z)$, when expanded in series, has only
non-negative powers of $z$, the integrand is finite and the
integral gives a well-defined operator in $\partial_{x_{\ell}}$
and $\partial_{u_{\ell}}$. One may similarly expand  $V^{\sst
(0)}(t)$ and $V^{\sst (1)}(t)$ \eqref{linear}  in powers of $t$,
and define the coefficients $V^{\sst (n)}_{m}$ for $n=0,1$ as \be
    V^{\sst (0)}_{m}=\delta_{m,0}\,,
    \qquad
    V^{\sst (1)}_{m}(\partial_{u})=\delta_{m\leqslant0}
    \left(\frac14\,\partial_{u}^{2}\right)^{\!-m}\,.
    \label{exp linear}
\ee
 Comparing this expansion with \eqref{K}, one can finally
obtain the HKE coefficients in terms of $V^{\sst (n)}_{m}$ as \ba
    a_{m}[g] \e \int \frac{d^{d}x}{(4\pi)^{\frac{d}{2}}}\,
    \sum_{n=0}^{\infty}\,\frac1n\,
    V^{\sst (n)}_{m-n}\!\left(\partial_{y_{1}},\frac{\partial_{u_{1}}+\partial_{u_{2}}}2;\cdots
    ;\partial_{y_{n}},\frac{\partial_{u_{n}}+\partial_{u_{1}}}2\right)\times \nn
    && \hspace{100pt} \times\,
    h(x_{1},u_{1})\,\cdots\,h(x_{n},u_{n})\,
    \Big|_{\overset{x_{1}=\cdots=x_{n}=x}{\sst u_{1}=\cdots=u_{n}=0}}\,.
    \label{int HKE}
\ea

\subsection{Linear and quadratic part of the heat kernel expansion coefficients}
\label{sec:quad heat}

Now we concentrate  on the
part of $\mathcal{K}[g|t]$ containing at most quadratic orders in
$h$\.: \ba
    \mathcal{K}[g|t] \e \int d^{d}x\ V^{\sst (0)}(t)+
    V^{\sst (1)}(\partial_{x},\partial_{u}|t)\,h(x,u)\,\Big|_{u=0}+\nn
    &&\qquad +\,\frac12\,V^{\sst (2)}(\partial_{x_{1}},\partial_{u_{1}};
    \partial_{x_{2}},\partial_{u_{2}}|t)\,
    h(x_{1},u_{1})\,h(x_{2},u_{2})\,\Big|_{\overset{x_{1}=x_{2}=x}{\sst u_{1}=u_{2}=0}}
    +\mathcal{O}(h^{3})\,.
\ea The constant and linear parts $V^{\sst (0)}, V^{\sst (1)}$ are
given in \eqref{linear}, and the quadratic part $V^{\sst (2)}$ is
obtained from  \eqref{Vn expansion} as a
series:
\be
    V^{\sst (2)}(\partial_{x_{1}},\partial_{u_{1}};
    \partial_{x_{2}},\partial_{u_{2}}|t)
    =(4\pi\,t)^{-\frac d 2}\,t^{2}
    \sum_{m=-\infty}^{\infty} t^{m}\,
    V^{\sst (2)}_{m}\!\left(\frac{\partial_{x_{12}}}2,\frac{\partial_{u_{12}}}2;
    -\frac{\partial_{x_{12}}}2,\frac{\partial_{u_{12}}}2\right)\,,
    \label{V2 expansion}
\ee
where $ \partial_{x_{12}}:=(\partial_{x_{1}}-\partial_{x_{2}})/2$
and $\partial_{u_{12}}:=\partial_{u_{1}}+\partial_{u_{2}}$\,,
and $V^{\sst (2)}_{m}$ is given from 
(\ref{Q n m} - \ref{ff}) by
\ba
    && V^{\sst (2)}_{m}(\partial_{x},\partial_{u}) :=
    V^{\sst (2)}_{m}\!\left(\frac{\partial_{x}}2,\frac{\partial_{u}}2;
    -\frac{\partial_{x}}2,\frac{\partial_{u}}2\right) \nn
    &&\quad =\,\left(\frac{\partial_{x}^{2}}{\partial_{u}^{2}}\right)^{\!\frac{m}2}
    \int_{-1}^{1}\frac{d\rho}2\,(1-\rho^{2})^{\frac m2}\,
    I_{m}\!\left( \frac12\sqrt{(1-\rho^{2})\,\partial_{x}^{2}\,\partial_{u}^{2}}\,\right)
    e^{i\,\frac{\rho}2\,\partial_{x}\cdot\partial_{u}}\,.
    \label{Q int}
\ea Even though the entire integrand is finite,
$(1-\rho^{2})^{m/2}$ superficially diverges for negative $m$, so
the integral should be treated separately for negative $m$. From
\eqref{int HKE} we see that $V^{\sst (2)}_{m}$ gives the
quadratic part of the HKE coefficient $a_{m+2}$. Notice that the case of 
negative $m$  corresponds to the HKE coefficients
$a_{n\leqslant1}$ where the linear terms \eqref{exp linear} appear.
Therefore, we will treat the
coefficients $a_{n\geqslant 2}$ and $a_{n\leqslant1}$ separately.

\subsubsection*{A. $a_{n\geqslant 2}$\,: HKE coefficients without linear term}

The integral \eqref{Q int} for $m\geqslant0$ can be directly evaluated
by using an integration formula of Bessel functions  recalled in
\eqref{int. Bessel} as \be
    V^{\sst (2)}_m(\partial_{x},\partial_{u})=
    \sqrt{\frac\pi2} \left(\frac12\,\partial_{x}^{2}\right)^{\!m}
    U_{m+\frac12}\!\left((\partial_{x}\cdot\partial_{u})^{2}-
    \partial_{x}^{2}\,\partial_{u}^{2}\right)
    \qquad [m\geqslant 0]\,,
    \label{n non-negative}
\ee
where $U_{\nu}$ is defined in \eqref{U bessel}.
By using this formula
we can obtain the explicit form of the HKE coefficients $a_{n\geqslant
2}$ up to quadratic term in $h$ as \ba
    a_{m+2}[g] \e \sqrt{\frac\pi8} \int \frac{d^{d}x}{(4\,\pi)^{\frac d2}}\,
    \left(\frac12\,\partial_{x_{12}}^{2}\right)^{\!m}
    U_{m+\frac12}\!\left((\partial_{x_{12}}\cdot\partial_{u_{12}})^{2}-
    \partial_{x_{12}}^{2}\,\partial_{u_{12}}^{2}\right)\times\nn
    && \hspace{80pt} \times\
    h(x_{1},u_{1})\,h(x_{2},u_{2})\,\Big|_{\overset{x_{1}=x_{2}=x}{\sst u_{1}=u_{2}=0}}
    +\mathcal{O}(h^{3})
    \qquad [m\geqslant0]\,.
    \label{quad inv a}
\ea Note that, as expected,  they are free from  constant or
linear terms in $h$. In other words, these are the lowest
$h$-order part of the HKE coefficients, and  as a consequence
they should be invariant under the lowest $h$-order part of
$\epsilon$-symmetry \eqref{nonlinear e transf}: $\delta^{\sst
[0]}_{\epsilon} h(x,u)=(u\cdot\partial_{x})\,\epsilon(x,u)$. This
gauge invariance can be checked from 
the identity: \be
    f\!\left((\partial_{x}\cdot\partial_{u})^{2}-
    \partial_{x}^{2}\,\partial_{u}^{2}\right)\,
    u\cdot\partial_{x}=
    u\cdot\partial_{x}\
    f\!\left((\partial_{x}\cdot\partial_{u})^{2}-
    \partial_{x}^{2}\,\partial_{u}^{2}\right),
\ee
satisfied by any function $f$\,:
the gradient $u_{1}\cdot\partial_{x_{1}}$ pass though 
$ U_{m+\frac12}\!\left((\partial_{x_{12}}\cdot\partial_{u_{12}})^{2}-
    \partial_{x_{12}}^{2}\,\partial_{u_{12}}^{2}\right)$, 
and the gauge variation vanishes when imposing $u_{1}=u_{2}=0$.

\medskip

Since the quadratic part of the HKE coefficients $a_{n\geqslant2}$ is
invariant under gauge transformations, it should be possible to
express it in terms of the higher-spin curvatures \eqref{HS curvature}. In
the rest of this subsection we show that this is indeed possible and
we give the expression. First, let us introduce a notation which will
be very convenient: when $v$ and $w$ are two vectors we denote by
$[vw]$ the antisymmetric matrix with elements
$[vw]^{\mu\nu}=v^{\mu}\,w^{\nu}-w^{\mu}\,v^{\nu}$\,, and $\left<\,
A\,\right>$ will be used  for the trace of the matrix $A$. Using
this notation we have  \be
    (v\cdot \partial_{x})\,(w\cdot\partial_{u})-(w\cdot\partial_{x})\,
    (v\cdot\partial_{u}) = \frac12\,\left<\,[vw]\,[\partial_{u}\partial_{x}]\,\right>,
    \qquad
    (\partial_{x}\cdot\partial_{u})^{2}-\partial_{x}^{2}\,\partial_{u}^{2}
    =\frac 12 \,\left<\,[\partial_{u}\partial_{x}]^{2}\right>.
\ee Now we make an ansatz for the quadratic part \eqref{quad inv
a} of $a_{m+2}$:  \ba
    && \sqrt{\frac\pi8} \int \frac{d^{d}x}{(4\,\pi)^{\frac d2}}\
    g_{m}([\partial_{v}\partial_{w}])\,
    R(x,v,w)\,\left(\frac12\,\partial_x^{2}\right)^{\!m}\,R(x,-v,w)\,\Big|_{v=w=0} \nn
    &&\quad =  \sqrt{\frac\pi8} \int \frac{d^{d}x}{(4\,\pi)^{\frac d2}}\,
    \left(\frac12\,\partial_{x_{12}}^{2}\right)^{\!m}\,
    \check g_{m}([\partial_{u_{12}}\partial_{x_{12}}])\,
     h(x_{1},u_{1})\,h(x_{2},u_{2})\,\Big|_{\overset{x_{1}=x_{2}=x}{\sst u_{1}=u_{2}=0}}\,,
     \quad  \label{ansatz}
\ea
where $R(x,v,w)$ is the generating function of higher-spin curvatures, 
$g_{m}$ a function which maps an antisymmetric matrix to a real number,
and the transformation $\check{(\cdot)}$ is defined by
\be
    \check{f}([xy]):=f([\partial_{v}\partial_{w}])\,
    e^{\frac12 \left<\, [vw]\,[xy]\,\right>}\,\Big|_{v=w=0}
    =\int_{0}^{\infty} dt\,t\,e^{-t}\,f(-t\,[xy])\,.
    \label{transf R}
\ee
The derivation of the second equality of the above equation is presented in
Appendix \ref{sec:curvature}.
By comparing \eqref{ansatz} to \eqref{quad inv a}, we get
$\check{g}_{m}([xy])=U_{m+\frac12}\!\left(\frac12\left<\,[xy]^{2}\right>\right)$
and finally
\be
    g_{m}([xy])=\sum_{n=0}^{\infty}\, \frac{(-1)^{n}\,2^{-(m+n+\frac12)}}
    {n!\,(2n+1)!\,\Gamma(m+n+\frac32)}
    \left(\frac{\left<\,[xy]^{2}\right>}{16}\right)^{\!n}\,.
\ee Thus, the quadratic part of the  coefficient $a_{n\geqslant 2}$
is expressed in terms of the generating function of higher-spin curvatures.

Using this expression for the HKE coefficients, we  now explicitly 
obtain,
up to the linear order in $h$,  the Weyl anomaly \eqref{Weyl anomaly} 
as
\be
	\mathscr{A}^{\sst(0)}[\bar h](x)= - \frac{N}{2\,(16\,\pi)^{\frac{d-1}2}}\,
	\sum_{n=0}^{\infty}\, \frac{1}{n!\,\Gamma(\frac{d+1}2+n)\,2^{4n}}\ 
	\Box^{\frac d2-1}_{x}\,R^{\sst (2n)}(x)+{\cal O}(h^2)\,,
\ee
where $R^{\sst (2n)}(x)$ is the linearized higher-spin scalar curvature,
that is the maximal trace of $R^{\sst(2n)}(x,v,w)$\,: 
\be
R^{\sst (2n)}(x):=
\left[\,\partial_{x}^{2}\,\partial_{u}^{2}-
(\partial_{x}\!\cdot\partial_{u})^{2}\,\right]^{n}\,h^{\sst (2n)}(x,u)\,.
\ee
Notice that only even higher-spin fields contribute to the result and that
the $n=1$ term reproduces    the linearized Weyl anomaly of gravity.
Indeed, the latter   is given in $d=2$ by the Ricci scalar $\mathcal{R}$\,, 
while for $d\geqslant 4$ it is given by
 $\Box_{x}^{d/2-1}\,\mathcal{R}$ plus 
other terms which are at least quadratic in  the fields (see \eqref{Weyl An} for the 
$d=4$ case).
 We  also express the generalized Weyl anomaly \eqref{HS Weyl anomaly} 
 in terms of higher-spin curvatures as
\ba
	&& \mathscr{A}^{\sst(r)}[\bar h](x,u)= - \frac{N}{2\,(16\,\pi)^{\frac{d-1}2}}\,\times\nn
	&& \qquad\qquad\times\,
	\sum_{n\geqslant \frac r2}^{\infty}\,
	\frac{(-1)^{n}\, \left<\,[\partial_{v}\partial_{w}]^{2}\,\right>^{n}
      \left<\,[vw]\,[u\partial_{x}]\,\right>^{r}}
      {r!\,n!\,(2n+1)!\,\Gamma(\frac{d+1}2+n)\,2^{r+5n}}\  
    \Box^{\frac d2-1}_{x}\,R^{\sst(2n-r)}(x,v,w) +{\cal O}(h^2)\,,\qquad
\ea 
and we notice the appearance once again of   
$\Box_{x}^{\frac d2 -1}$\,, but now it acts on the  traces 
of the linearized higher-spin curvatures
which have $r$ free indices. The linearized trace anomaly obtained from the quadratic part of the effective action of a conformally coupled scalar field on $AdS_4$ in the presence of a single external higher-spin gauge field was considered in \cite{Manvelyan:2005ew}.

Now we come back to the expression \eqref{ansatz}
and notice that when decomposed  
in spin components 
we  get  couplings between curvatures of different spins. 
Indeed, the curvatures that we have introduced are
associated to $h(x,u)$ and one may wonder whether the HKE coefficients
get diagonalized in terms of $\bar h$.
In order to see that, it will be useful to first express
$U_{m+\frac12}\!\left(\frac12 \left<\,[
\partial_{u_{12}}\partial_{x_{12}}]^{2}\right>\right)$,
by making use of the
addition theorem of Bessel functions, as
\ba
    && U_{m+\frac12}\!\left(\frac12 \left<\,
    [ \partial_{u_{12}}\partial_{x_{12}}]^{2}\, \right> \right)
    =2^{m+\frac12}\,\Gamma\!\left(m+\frac12\right)\times
    \nn
    &&\times\,\sum_{s=0}^{\infty}
    \left(s+m+\frac12\right)
    \left(\sqrt{\left<\,[ \partial_{u_{1}}\partial_{x_{12}}]^{2}\right>\!
    \left<\,[ \partial_{u_{2}}\partial_{x_{12}}]^{2}\right>}\right)^{s}
    C_{s}^{m+\frac12}\!\left(\frac{
    \left<\,[ \partial_{u_{1}}\partial_{x_{12}}][ \partial_{u_{2}}\partial_{x_{12}}]\,
    \right>}{\sqrt{
    \left<\,[ \partial_{u_{1}}\partial_{x_{12}}]^{2}\right>\!
    \left<\,[ \partial_{u_{2}}\partial_{x_{12}}]^{2}\right>}} \right)\times \nn
    && \qquad\qquad \times\,
     U_{s+m+\frac12}\!\left(\frac12 \left<\,
    [ \partial_{u_{1}}\partial_{x_{12}}]^{2}\, \right> \right)
    U_{s+m+\frac12}\!\left(\frac12 \left<\,
    [ \partial_{u_{2}}\partial_{x_{12}}]^{2}\, \right> \right),
\ea where $C_{s}^{\lambda}(z)$ is the Gegenbauer polynomial. Since
$\omega^{s}\,C_{s}^{\lambda}(z/\omega)$ is a polynomial of order
$s$ in $z$ and $\omega$, when contracting the above with $h$ and
integrating by part, we will pick the  homogeneous term of order
$s$  in $U_{s+m+\frac12}\!\left(\frac12 \left<\,
    [ \partial_{u}\partial_{x}]^{2}\, \right> \right)\,h(x,u)$
in the summation:
\be
    \left(U_{s+m+\frac12}\!\left(\frac12 \left<\,
    [ \partial_{u}\partial_{x}]^{2}\, \right> \right)\,h\,\right)^{\!\!\sst(s)}\!\!(x,u)
    =\frac{ 2^{-(s+m+\frac12)}}{\Gamma(s+m+\frac32)}
    \left(\Pi_{2m+4}^{-1}(i\,\partial_{u},\partial_{x})\,h\right)\!{}^{\sst (s)}(x,u)\,,
\ee
where we have used \eqref{Pi'}.
Finally the quadratic part of $a_{m+2}$ can be written as
\ba
    && \sqrt{\frac\pi8} \int \frac{d^{d}x}{(4\,\pi)^{\frac d2}}\,
    \left(\frac12\,\partial_{x_{12}}^{2}\right)^{\!m}
    G_{m}\Big( \left<\,[ \partial_{u_{1}}\partial_{x_{1}}][ \partial_{u_{2}}\partial_{x_{2}}]\,
    \right>, \left<\,[ \partial_{u_{1}}\partial_{x_{1}}]^{2}\right>\!
    \left<\,[ \partial_{u_{2}}\partial_{x_{2}}]^{2}\right> \Big)\,\times \nn
    && \quad \times \,\Pi_{2m+4}^{-1}(i\,\partial_{u_{1}},\partial_{x_{1}})\,
    \Pi_{d}(i\,\partial_{u_{1}},\partial_{x_{1}})\,
    \bar h(x_{1},u_{1})\
    \Pi_{2m+4}^{-1}(i\,\partial_{u_{2}},\partial_{x_{2}})\,
     \Pi_{d}(i\,\partial_{u_{2}},\partial_{x_{2}})\
    \bar h(x_{2},u_{2})\,
    \Big|_{\overset{\sst x_{1}=x_{2}=x}{\sst u_{1}=u_{2}=0}}\,,
    \nonumber
    \label{quad a}
\ea
with
\be
    G_{m}(z,\omega)=\sum_{s=0}^{\infty}
     \frac{2^{-(2s+m+\frac12)}\, \Gamma(m+\frac12)}
     {\Gamma(s+m+\frac12)\,\Gamma(s+m+\frac32)}\,
    \omega^{\frac s2}\,C^{m+\frac12}_{s}\!\left(\frac{z}{\sqrt{\omega}}\right).
    \label{quad a ker}
\ee Here one can see that when $m=(d-4)/2$, the HKE coefficient
is diagonalized in terms of $\bar h$, but in general it is not the
case. One might consider new current generator
$J_{m}(x,q):=\Pi_{2m+4}(q,\partial_{x})\,J(x,q)$ and couple them to new 
external higher-spin fields
$h_{m}(x,u)$, then the quadratic part of HKE coefficient
$a_{m+2}$ will be diagonal in $h_{m}(x,u)$ and is given by
\ba
    && \sqrt{\frac\pi8} \int \frac{d^{d}x}{(4\,\pi)^{\frac d2}}\,
    \left(\frac12\,\partial_{x_{12}}^{2}\right)^{\!m}
    F_{m}\Big( \left<\,[ \partial_{v_{1}}\partial_{w_{1}}][ \partial_{v_{2}}\partial_{w_{2}}]\,
    \right>, \left<\,[ \partial_{v_{1}}\partial_{w_{1}}]^{2}\right>\!
    \left<\,[ \partial_{v_{2}}\partial_{w_{2}}]^{2}\right> \Big)\,\times \nn
    && \hspace{120pt} \times\, R_{m}(x_{1},v_{1},w_{1})\
   R_{m}(x_{2},v_{2},w_{2})\,
    \Big|_{\overset{\sst x_{1}=x_{2}=x}{\sst v_{1}=v_{2}=w_{1}=w_{2}=0}}\,,
\label{courb}
\ea
where $R_{m}$ is the higher-spin curvature associated to $h_{m}$, and
$F_{m}$ can be obtained again by using \eqref{transf R} as
\be
    F_{m}(z,\omega)=\sum_{s=0}^{\infty}
     \frac{2^{-(2s+m+\frac12)}\, \Gamma(m+\frac12)}
     {\Gamma^{2}(s+2)\,\Gamma(s+m+\frac12)\,\Gamma(s+m+\frac32)}\,
    \omega^{\frac s2}\,C^{m+\frac12}_{s}\!\left(\frac{z}{\sqrt{\omega}}\right),
\ee
but all the other HKE
coefficients will  remain un-diagonalized.

\subsubsection*{B. $a_{d/2}$\,: Weyl invariant HKE coefficient}

Among all the HKE coefficients, of particular interest is
$a_{d/2}$ which exists only for even dimension and admits the higher-spin
Weyl symmetry \eqref{nonlinear a transf}\,. For $d\geqslant4$, 
the quadratic
part of $a_{d/2}$ is given by \ba
    a_{\frac d2}[\bar h]\e  \sqrt{\frac\pi8}
    \int \frac{d^{d}x}{(4\,\pi)^{\frac {d}2}}\,
    G_{\frac{d-4}2}\Big( \left<\,[ \partial_{u_{1}}\partial_{x_{1}}][ \partial_{u_{2}}\partial_{x_{2}}]\,
    \right>, \left<\,[ \partial_{u_{1}}\partial_{x_{1}}]^{2}\right>\!
    \left<\,[ \partial_{u_{2}}\partial_{x_{2}}]^{2}\right> \Big)\,\times \nn
    && \qquad \times
    \left(\frac12\,\partial_{x_{12}}^{2}\right)^{\!\frac {d-4}2 }\,
    \bar h(x_{1},u_{1})\,\bar h(x_{2},u_{2})\,
    \Big|_{\overset{\sst x_{1}=x_{2}=x}{\sst u_{1}=u_{2}=0}}\,
    +\mathcal{O}(\bar h^{3})\,,
\ea and  it coincides with the result found in \cite{Segal:2002gd}. 
Its expression in terms of curvatures of $\bar h$
is obtained from  (\ref{courb}) by replacing $R_m$ with $\bar R$ and $m$ with 
$(d-4)/2$.

The quadratic part of $a_{d/2}$ is invariant under the Abelian part of 
the generalized Weyl transformation:
$\delta^{\sst [0]}_{\bar\alpha} \bar h(x,u) = u^{2}\,\bar\alpha(x,u)$. 
This can be checked 
by computing
\be
    G_{m}\Big( \left<\,[ \partial_{u_{1}}\partial_{x_{1}}][ \partial_{u_{2}}\partial_{x_{2}}]\,
    \right>, \left<\,[ \partial_{u_{1}}\partial_{x_{1}}]^{2}\right>\!
    \left<\,[ \partial_{u_{2}}\partial_{x_{2}}]^{2}\right> \Big)\,u_{1}^{2}\,
    \bar\alpha(x_{1},u_{1})\,\Big|_{u_{1}=0}\,,
\ee
which is simplified for $m=(d-4)/2$ thanks to the differential equation \eqref{de Gen}
of the Gegenbauer polynomial. The latter  allows to factor out in the above expression 
 the  operator 
\be
    \left<\, [\partial_{u_{2}}\partial_{x_{2}}]\,[\partial_{x_{2}}\partial_{x_{1}}]\,
    [\partial_{x_{1}}\partial_{u_{2}}]\,\right>,
\ee
which gives a total derivative term.
In fact, the generalized Weyl invariance can be more easily checked
with the undiagonalized formula \eqref{quad inv a}.
By using
\ba
    && U_{m+\frac12}\!\left((\partial_{x}\cdot\partial_{u})^{2}-
    \partial_{x}^{2}\,\partial_{u}^{2}\right)
    \left(u^{2}-\frac14\,\partial_{x}^{2}\right)=
    u^{2}\,U_{m+\frac12}\!\left((\partial_{x}\cdot\partial_{u})^{2}-
    \partial_{x}^{2}\,\partial_{u}^{2}\right)+\nn
    &&\quad +\, \frac12\left(
    u\cdot\partial_{u}+
    \frac{d-4}2-m-\frac{(u\cdot\partial_{x})\,(\partial_{x}\cdot\partial_{u})}
    {\partial_{x}^{2}}\right)
    U_{m+\frac32}\!\left((\partial_{x}\cdot\partial_{u})^{2}-
    \partial_{x}^{2}\,\partial_{u}^{2}\right)\,,
    \label{Weyl inv fn}
\ea
with  $m=(d-4)/2$, the invariance of \eqref{quad inv a}
under $\delta^{\sst [0]}_{\alpha} h(x,u) = (u^{2}-\partial_{x}^{2}/4)\,\alpha(x,u)$
is easily shown.

The higher-spin Weyl invariance of the above quadratic term together with the
number of derivatives involved implies that it can be simply
expressed in terms of the higher-spin Weyl tensor. The latter is the traceless
part of the curvature tensor and belongs to the same Young tableau
representation.
Since the unique Fronsdal and Weyl (like) invariant expression with $2s$
derivatives is the square of the higher-spin Weyl tensor, we conclude that
the HKE coefficient density
$b_{d/2}$ is proportional to the sum over all spins of the
squares of the corresponding higher-spin Weyl tensors, up to a total derivative term. 
This is the free action
considered by \xav{Fradkin, Tseytlin and Segal \cite{Fradkin:1985am,Segal:2002gd}.}

\subsubsection*{C. $a_{n\leqslant1}$\,: HKE coefficients with linear terms}

Finally we compute the HKE coefficients $a_{n\leqslant1}$ up to
quadratic order in the fields and get \ba
    a_{1-m}[g] \e \int \frac{d^{d}x}{(4\,\pi)^{\frac d2}}\ \
    \delta_{m,1}+ \left(\frac14\,\partial_{u}^{2}\right)^{m}
    h(x,u)\,\Big|_{u=0}+ \nn
    &&+\,\frac12\,V^{\sst (2)}_{-(m+1)}(\partial_{x_{12}},\partial_{u_{12}})\,
    h(x_{1},u_{1})\,h(x_{2},u_{2})\,\Big|_{\overset{x_{1}=x_{2}=x}{\sst u_{1}=u_{2}=0}}
    +\mathcal{O}(h^{3})\qquad [m\geqslant 0]\,,\qquad
    \label{quad non inv a}
\ea where the integral \eqref{Q int} for $V^{\sst (2)}_{-(m+1)}$
can be evaluated by expanding the Bessel function  and we get
\be
    V^{\sst (2)}_{-(m+1)}(\partial_{x},\partial_{u})=
    \sqrt{\frac\pi2}\,\left(\frac14\,\partial_{u}^{2}\right)^{\!m+1}
    \sum_{k=0}^{\infty}\,\frac{
    \left(\frac18\,\partial_{x}^{2}\,\partial_{u}^{2}\right)^{\!k}
    }{\Gamma(k+m+2)}
    \left(\frac{\partial_{x}\cdot\partial_{u}}2\right)^{\!-k-\frac12}\!
    J_{k+\frac12}\!\left(\frac{\partial_{x}\cdot\partial_{u}}2\right).
    \label{V neg}
\ee Since $z^{-\nu}\,J_{\nu}(z)$ has an expansion with only
non-negative integer powers of $z^{2}$, $V^{\sst (2)}_{-(m+1)}$ is
a well defined operator containing at least the $(m+1)$-th power
of $\partial_{u}^{2}$. Due to  these traces, $V^{\sst
(2)}_{-(m+1)}$ does not commute with $u\cdot\partial_{x}$ and the
coefficient is not invariant under the free gauge transformation.
This is expected from the presence of the linear term: the linear
variation of the quadratic part must be cancelled by the quadratic
variation of linear part. One may wonder whether there exists a
natural decomposition of $V^{\sst (2)}_{-(m+1)}$ into a gauge
invariant part and a gauge non-invariant part which compensates
the variation of  the linear part. In order to do so we use the
Lommel expansion \eqref{Lommel} to evaluate the infinite series
sum by adding $m+1$ terms as \ba
     && V^{\sst (2)}_{-(m+1)}(\partial_{x},\partial_{u}) =
     \sqrt{\frac\pi2} \left(\frac12\,\partial_{x}^{2}\right)^{\!-m-1}\times\nn
     && \times\,\left[
     U_{-m-\frac12}\!\left((\partial_{x}\cdot\partial_{u})^{2}-
     \partial_{x}^{2}\,\partial_{u}^{2}\right) -\sum_{k=0}^{m}\,\frac{
    \left(\frac18\,\partial_{x}^{2}\,\partial_{u}^{2}\right)^{\!k}
    }{k!}
    \left(\frac{\partial_{x}\cdot\partial_{u}}2\right)^{\!-(k-m-\frac12)}\!
    J_{k-m-\frac12}\!\left(\frac{\partial_{x}\cdot\partial_{u}}2\right)\right].\nn
    \label{V neg sep}
\ea Notice that the first part of second line coincides with the
expression \eqref{n non-negative} for $V^{\sst(2)}_{n\geqslant 0}$ with
$n=-m-1$ and thus is gauge invariant, while the second part, that
is the finite series with $m+1$ terms, is not gauge invariant. The
price to pay for separating $V_{-(m+1)}^{\sst (2)}$ in this way is
the locality: the initial expression \eqref{V neg} is  local but
if we rewrite it as \eqref{V neg sep} then each gauge invariant or
gauge non-invariant part becomes non-local, or in other words the
non-local terms of each part cancel out.

\subsection{Quadratic part of the renormalized effective action}
\label{sec:finite}

The regularized effective action $\mathcal{W}_{\rm reg}[\bar
h|\Lambda]$ has an expansion \eqref{EA expansion} in
$\Lambda$. All the terms  except the finite part $\mathcal{W}_{\rm
fin}[\bar h|\mu]$  are directly given by the HKE coefficients. In
this section we will compute the remaining term
$\mathcal{W}_{\rm fin}[\bar h]$ up to the quadratic order in $h$.
It can be obtained as  
\be
\mathcal{W}_{\rm fin}[\bar h]=\lim_{\Lambda\rightarrow \infty}
\Big(\, \mathcal{W}_{\rm reg}[\bar
h|\Lambda]-\mathcal{W}_{\rm div}[\bar
h|\Lambda]\,\Big)\,,
\ee
where $\mathcal{W}_{\rm div}[\bar
h|\Lambda]$ is the divergent part of the effective action.
From equations  \eqref{reg. schwinger} and \eqref{EA expansion}
we see that the finite part  of the effective action receives contributions 
only  from non-negative powers of $t$
in $\mathcal{K}[g|t]$\,:
\be
     t^{-\frac{d}2}\, \sum_{n\geqslant \frac {d}2}^{\infty}\,
    t^{n}\,a_{n}[g]\,,
\ee
 that is, from the HKE coefficients $a_{n}[g]$
with $n\geqslant d/2$\,.

In fact, for $d>2$, it is more convenient to first replace $\mathcal{K}[g|t]$
in \eqref{reg. schwinger}
by
\be
\mathcal{K}_{\sst \geqslant2}[g|t]=t^{2-\frac{d}2} \sum_{m=0}^{\infty}
    t^{m}\,a_{m+2}[g]\,,
\ee
because, as we will see below, the above expression can be
exactly resummed at the quadratic order in $h$.
Next we  calculate the integral over $t$ and  
then subtract the divergent terms. 
Indeed, using the expression of the quadratic part of $a_n[g]$ \eqref{quad inv a}
we get
\be
    \mathcal{K}_{\sst \geqslant2}[g|t]= \int \frac{d^{d}p}{(2\pi)^{d}}\,
    V^{\sst (2)}_{\sst \geqslant 2}(p,\partial_{u_{12}}|t)\
    \tilde h(-p,u_{1})\,\tilde h(p,u_{2})\,\Big|_{u_{1}=u_{2}=0}
    + \mathcal{O}(h^{3})\,,
\ee
with $\partial_{u_{12}}:=\partial_{u_{1}}+\partial_{u_{2}}$ and
\be
V^{\sst (2)}_{\sst \geqslant 2}(p,\partial_{u}|t):=
(4\pi\,t)^{-\frac d2}\,\sum_{m=0}^\infty\,
\sqrt{\frac{\pi}8}\,\left(-\,\frac{p^{2}}2\,t\right)^{\!m}\,
U_{m+\frac12}\big(p^2\,\partial_{u}^2-(p\cdot\partial_{u})^2\big)\,.
\label{V>=2}
\ee
The
expansion in powers of $t$ is not convenient because positive powers of $t$ are not
integrable separately. As we mentioned before we can resum \eqref{V>=2}
in $t$
by using the series representation of Bessel functions to get
\be
    V^{\sst (2)}_{\sst \geqslant2}
    (p,\partial_{u}|t)
    =(4\pi\,t)^{-\frac d2}\,t^{2}\,\sum_{m=0}^{\infty}\,\frac{
    \left[(p\cdot\partial_{u})^{2}-{p}^{2}\,\partial_{u}^{2}
    \right]^{m}}{(2m+1)!\,2^{2m}}\,
    {}_{1}F_{1}\!\left(1;m+\frac32;-\,\frac{p^{2}}4\,t\right)\,.
\ee
We then integrate the hypergeometric function 
and subtract the divergent parts
as explained in Appendix \ref{sec:int. 1F1},
and finally we obtain the finite part of the effective action.
 For odd $d$\,, it is given by \ba
    && \mathcal{W}_{\rm fin}[\bar h] = N\,(-1)^{\frac{d-1}2}\pi\,
    \frac{(2\pi)^{-\frac{d-1}2}}{8}\,
    \int \frac{d^{d} p}{(2\pi)^{d}}
    \left(\frac{p^{2}}4\right)^{\frac{d-4}2}\,\times\nn
    && \quad \times\,
    U_{\frac{d-3}2}
    \left( p^{2}\,(\partial_{u_{1}}+\partial_{u_{2}})^{2}
    -[p\cdot(\partial_{u_{1}}+\partial_{u_{2}})]^{2} \right)\,
    \tilde h(-p,u_{1})\ \tilde h(p,u_{2})\,\Big|_{\sst u_{\ell}=0}
    \!\!+\mathcal{O}(h^{3})\,,\quad
    \label{W fin odd d}
\ea and for even $d>2 $,  by \ba
    && \mathcal{W}_{\rm fin}[\bar h|\mu] = N\,(-1)^{\frac d2}\,
    \frac{(2\pi)^{-\frac{d-1}2}}{8}\,
    \int \frac{d^{d} p}{(2\pi)^{d}}
    \left(\frac{p^{2}}4\right)^{\frac{d-4}2}\,\times\nn
    && \times\,\bigg[
    \ln\!\left(\frac{p^{2}}{\mu^{2}}\right)\,
    U_{\frac{d-3}2}
    \left( p^{2}\,(\partial_{u_{1}}+\partial_{u_{2}})^{2}
    -[p\cdot(\partial_{u_{1}}+\partial_{u_{2}})]^{2}\right) +\nn
    && \quad +\,
    \dot{U}_{\frac{d-3}2}
    \left( p^{2}\,(\partial_{u_{1}}+\partial_{u_{2}})^{2}
    -[p\cdot(\partial_{u_{1}}+\partial_{u_{2}})]^{2} \right)\bigg]\
    \tilde h(-p,u_{1})\ \tilde h(p,u_{2})\,\bigg|_{\sst u_{\ell}=0}
    \!\!+\mathcal{O}(h^{3})\,,\qquad
    \label{W fin even d}
\ea where $\dot{U}_{\nu}(z):= (\partial/
\partial\nu)\,U_{\nu}(z)$\,
and we introduced  a mass scale  $\mu$  in order to make the
argument of $\ln$ dimensionless.

 The $d=2$ case
 requires two additional considerations. First one should
take $a_{1}$ into account, and second the integral over $t$ \eqref{reg. schwinger}
should also be
regularized  in the infrared.  A convenient regulator is provided
by inserting  $(\nu^{2}\,t)^{-\xi}$ with $0<\xi \ll 1$ into the
integration \eqref{reg. schwinger} over $t$. A constant $\nu$ with mass dimension is introduced
in order to make the regulator dimensionless.
The contribution of the $a_{n \geqslant 2}$ terms  can be calculated as before and
results in an expression which is given by the right hand side of 
 of \eqref{W fin even
d} with $d=2$ and where the ultraviolet regularization
ambiguity $\mu$ is replaced by the infrared regularization
ambiguity $\nu$\,. This expression will be denoted 
$\mathcal{W}_{\geqslant 2}[\bar h|\nu]$.
 The 
  $a_{1}$ contribution is both
 UV and IR divergent,
and its regularization yields
$2\,N\,\ln(\nu/\mu)\, a_{1}$\,. 
Finally the finite part of the  effective action for $d=2$ is given by
 \be
    \mathcal{W}_{\rm fin}[\bar h|\mu,\nu]=
    \mathcal{W}_{\geqslant 2}[\bar h|\nu]+
    2\,N\,\ln(\nu/\mu)\, a_{1}[g] +\mathcal{O}(h^{3})\,.
\ee

Several remarks are in order. 
\begin{itemize}
\item 
As is shown in \eqref{Weyl inv fn},
the terms expressed in terms of
$U_{(d-3)/2}$ are invariant under the linearized higher-spin Weyl transformation.
For odd $d$ we can see that the finite part
(which is nonlocal due to the half integer power of $p^{2}$) of
the regularized effective action is not anomalous, while for even
$d$ the variation of $\mathcal{W}_{\rm fin}[\bar h|\mu]$ does not
vanish due to the terms expressed in terms of the function
$\dot{U}_{(d-3)/2}$.
\item
The anomaly-free term expressed in terms of ${U}_{(d-3)/2}$ is diagonalized when
expressed in terms of $\bar h$ while the anomalous term with
$\dot{U}_{(d-3)/2}$ cannot be diagonalized and so results in
couplings of different spin fields at the quadratic level.
\item
The quadratic anomalous term with $\dot{U}_{(d-3)/2}$ in even $d$ 
can be expressed, by applying \eqref{ansatz},  in terms of the linearized higher-spin curvatures as 
\be
	\mathcal{W}_{\rm fin}^{A}[\bar h]
	=N\,
    \int {d^{d} x}\ f\!\left(\left<[\partial_{v}\partial_{w}]^{2}\right>\right)\,
    R(x,v,-w)\,\Box_{x}^{\frac{d-4}2}\,R(x,v,w)\,\Big|_{v=w=0}
    +\mathcal{O}(h^{3})\,,\label{nom}
\ee
where the function $f$ is given by the transformation \eqref{transf R}
of $\dot U_{(d-3)/2}$. 
While the  part of the renormalized effective
action which is invariant under higher-spin Weyl transformations always contains a non-local term, the quadratic anomalous
term in even $d$ is local, except for $d=2$ where the power of $\Box_{x}$ becomes negative. In $d>2$, the anomalous term may be compensated by a local counter-term (as all divergent parts of
the effective action). 
The exceptional $d=2$ anomaly
source term cannot be compensated by a local counter-term.
\item
Let us compare our results on higher-spins with standard gravity, \textit{i.e.} a complex scalar field 
in a curved spacetime background. 
The Weyl anomaly is not present for odd $d$. For $d$ equal to $2$ or $4$, it is given by
\cite{Birrell:1982ix}
\be
	\mathscr{A}^{\sst (0)}[g_{\mu\nu}]\propto\left\{
	\begin{array}{ccc}
	& \mathcal{R} \qquad & [d=2] \\
	&\Box_{x}\, \mathcal{R}
	+\mathcal{R}_{\mu\nu}^{2}-\frac13\,\mathcal{R}^{2}+
	 \mathcal{C}_{\mu\nu\rho\sigma}^{2}
	 \qquad &[d=4]
	\end{array}
	\right.
	\label{Weyl An}\,,
\ee
where 
$\mathcal{R}_{\mu\nu}$, $\mathcal{R}$ 
and $\mathcal{C}_{\mu\nu\rho\sigma}$ are 
respectively  the Ricci tensor, the Ricci scalar
and the Weyl tensor. 
The $d=2$ Weyl anomaly is reproduced by varying \cite{Polyakov:1981rd}
\be
	\mathcal{W}_{\rm fin}^{A}[g_{\mu\nu}]
	\propto\int d^{2}x\, \sqrt{g}\, \mathcal{R}\,\Box_{x}^{-1}\, \mathcal{R} \qquad [d=2]\,,
\ee
which is analogous to our finite part of the effective action \eqref{nom} with $d=2$.	
 The $\Box_x\,\mathcal{R}$ contribution to the four dimensional anomaly can be compensated by the
local  counter-term\,:
\be
	\mathcal{W}^{A \& {\rm local}}_{\rm fin}[g_{\mu\nu}]
	\propto\int d^{d}x\,  \sqrt{g}\,\mathcal{R}^{2} \qquad [d=4]\,,
\ee
which is also present in \eqref{nom} with $d=4$.
The quadratic terms in the anomaly can be obtained by varying   
terms of order $h^3$ in the effective
action which we have not explicitly calculated.

\item
Finally, the terms inside the kinetic operators of (\ref{W fin odd d}) and (\ref{W fin even d}) expressed via $U_{(d-3)/2}$ are accompanied with either
a half-integer power of $p^2$ (for odd $d$) or the $\log$ of $p^{2}$ (for even $d$) which are non-local operators.
For $d>2$, both of them are the Fourier transform of the Hadamard finite part of 
$1/(x^{2})^{d-2}$ given by
\be
-
    \frac{\pi^{\frac d2}\left(\frac{p^{2}}4\right)^{\frac{d-4}2}}
    {\Gamma(d-2)\,\Gamma(\frac{d-2}2)}\,
    \left\{
    \begin{array}{cc}
    (-1)^{\frac{d-1}2}\,\pi
     \qquad & [\,d\in 2\,\mathbb{N}+1\,]
      \vspace{5pt} \\
     (-1)^{\frac d2}\,\ln\!\left(\frac{p^{2}}{\mu^{2}}\right)      \qquad
    & [\,d\in 2\,\mathbb{N}+2\,]
    \end{array}\right..\quad
\ee
In the next section,
 we will see that this corresponds in fact to the two-point correlation function of the conserved current generator $J(x,q)$.


\end{itemize}

\subsection{Correlation functions}
\label{sec:correl}

The correlation functions of the currents can be obtained from the
effective action by calculating the functional derivatives with
respect to the external higher-spin fields. A simpler method is to
consider the correlation function of the generating function:
\be C_{n}(x_{1},q_{1};\cdots;x_{n},q_{n})=\left<\,J(x_1,q_1)\,\cdots
    J(x_n,q_n)\,\right>_{\rm connected}\,,\ee and  calculate it using
    the two-point function:
    \be
    \Delta(x_1-x_2)=\left<\, \phi(x_1)^*\phi(x_2)\, \right>=\int \frac{d^d p}{(2\pi)^d}\,
    \frac{e^{i\,p\cdot[x_1-x_2]}}{p^{2}}=
    \frac{2^{-2}\,\pi^{-\frac d2}\,\Gamma(\frac{d-2}{2}) }{\left[(x_1-x_2)^{2}\right]^{\frac{d-2}2}}\qquad [d>2]\,.
    \label{I}
    \ee
    Wick's theorem  and an expansion in the auxiliary variables allow to get
    the
    correlation functions of the currents. The n-point function of
    the generating function is thus given by
    \ba
   \label{decompo I}
    && C_{n}(x_{1},q_{1};\cdots;x_{n},q_{n}):=\left<\,J(x_1,q_1)\,\cdots
    J(x_n,q_n)\,\right>_{\rm connected}=\\
    && \ =\,\frac{N}{n\,n!}
    \sum_{\sigma\in \mathfrak{S}_n}
    \Delta\!\left(x_{\sigma_1}-x_{\sigma_2}+\tfrac{q_{\sigma_1}+q_{\sigma_2}}2\right)
    \Delta\!\left(x_{\sigma_2}-x_{\sigma_3}+\tfrac{q_{\sigma_2}+q_{\sigma_3}}2\right)
    \cdots\,\Delta\!\left(x_{\sigma_n}-x_{\sigma_1}+\tfrac{q_{\sigma_n}+q_{\sigma_1}}2\right),\nonumber
\ea
where $\mathfrak{S}_n$ is the symmetric group of order $n$.
    Let us illustrate this with the simplest example of the
    two-point functions:
\ba
    && C_{2}(x_{1},q_{1};x_{2},q_{2})
    =\left<\,J(x_1,q_1)\,J(x_2,q_2)\,\right>_{\rm connected}= \nn
    && \qquad =\,N\,
    \frac{2^{-5}\,\pi^{-d}\,\left[\Gamma(\frac{d-2}{2})\right]^{2}}
    {\left[\left\{(x_{1}-x_{2})^{2}+(\frac{q_{1}+q_{2}}2)^{2}\right\}^{2}
    -\big\{(x_{1}-x_{2})\cdot(q_{1}+q_{2})\big\}^{2}\right]^{\frac{d-2}2}}\,.
\ea   When expanded in $q$, $C_{2}$ reproduces the non-local part
of \eqref{W fin odd d} and \eqref{W fin even d} thanks to the
following identity: \be
    \frac1{\left[ (x^{2}+q^{2})^{2}-4\,(x\cdot q)^{2}\right]^{\frac{d-2}2}}
    =U_{\frac{d-3}2}\!\left(-(q\cdot\partial_{x})^{2}+
    q^{2}\,\partial_{x}^{2}\right)\,
    \frac{2^{\frac{d-3}2}\,\Gamma\!\left(\frac{d-1}2\right)}{\left(x^{2}\right)^{d-2}}\,,
    \label{corr Bessel}
\ee whose derivation  is presented in Appendix
\ref{sec:correlation}.

\section{Free scalar fields in the ordinary lower spin fields background}
\label{sec:low}

In the previous sections, we have considered the effective action
of scalar fields coupled to higher-spin fields contained in the
generating function $h(x,u)$, where the couplings are always
linear in the external fields. In fact, more generally, $h(x,u)$
can also be considered as composites of some other external fields
$\varphi(x,u)$. Such an example was briefly mentioned in the
introduction, it consists in replacing the scalar field in
$h(x,u)$ by the solution to $\mathcal{W}_{-1}[h]=0$ (or $a_{1}[g]=0$).
 So the
most general quadratic action of a (complex) scalar field $\chi$
in the background of external fields $\varphi$ can be written as
\be
    \mathscr{S}[\chi; \varphi]:=\bra{\phi_{\chi,\varphi}} \hat P^{2}-\hat H_{\varphi}
    +m^{2} \ket{\phi_{\chi,\varphi}}\,,
    \label{S H}
\ee
with some proper redefinitions of fields $\phi_{\chi,\varphi}(x)=f[\varphi](x)\,\chi(x)$
and $h_{\varphi}(x,u)=h_{\varphi}[\varphi](x,u)$.
Since the above has essentially the same form as \eqref{S class},
its effective action can be also regularized and computed in the same way as
\be
    \mathscr{W}[\varphi|\Lambda] := -
    \int_{\frac1{\Lambda^{2}}}^{\infty} \frac{dt}t\,e^{-t\,m^{2}}\,
    \mathcal{K}[\,p^{2}-h_{\varphi}(x,p)\,|\,t\,]\,,
\ee where $\mathcal{K}[\,p^{2}-h_{\varphi}(x,p)\,|\,t\,]$ is the
trace of the heat kernel which was defined in \eqref{heat trace
with A}.

In the present section, we consider the ordinary lower spin fields interactions,
that is, the electromagnetic and the gravitational interactions,
and compute the corresponding HKE coefficients with which
the effective action can be obtained.
It is well-known that their effective actions provide infinitely many gauge invariant actions
at each order of the cut-off scale.
Among them, one can find in particular the Maxwell action and the Hilbert-Einstein action.
In other words, the heat kernel of the Laplace equation properly dressed in $A_{\mu}$
or $g_{\mu\nu}$ give, as its expansion coefficients, infinitely many gauge invariants of
these fields.
In the followings, we show how to obtain the gauge invariants of
$A_{\mu}$ and $g_{\mu\nu}$ by applying the results of this letter.

\subsection{Electromagnetic interaction}

The free complex scalar field action in the electromagnetic background:
\be
    \int d^{d}x\,
    \left(\partial^{\mu}+i\,A^{\mu}\right)\chi^{*}
     \left(\partial_{\mu}-i\,A_{\mu}\right)\chi
    +m^{2}\,\chi^{*}\,\chi\,,
\ee
can be written as \eqref{S H} with $\phi_{\chi,A_{\mu}}=\chi$ and
\be
    \hat H_{A_{\mu}}= -A_{\mu}(\hat X)\,A^{\mu}(\hat X)-
    \left(A^{\mu}(\hat X)\,\hat P_{\mu}+\hat P_{\mu}\,A^{\mu}(\hat X)\right).
\ee By using the Wigner map, we identify the components of
$h_{A_{\mu}}(x,u)$ as \be
    h^{\sst (0)}_{A_{\mu}}(x)=-A_{\mu}(x)\,A^{\mu}(x)\,,
    \qquad
    h^{\sst (1)}_{A_{\mu}}(x,u)= -2\,u^{\mu}\,A_{\mu}(x)\,,
    \qquad
    h^{\sst (n\geqslant2)}_{A_{\mu}}(x,u)=0\,.
\ee Notice that the non-linear gauge symmetry \eqref{nonlinear e
transf} of $h_{A_{\mu}}$ defines the usual linear gauge symmetry
of the Maxwell fields. Notice also that   $h^{\sst (0)}_{A_{\mu}}$
and $h^{\sst (1)}_{A_{\mu}}$ are of different orders in $A_\mu$.

The HKE coefficients $a_{n\geqslant 2}$ can be obtained up to the quadratic order in $A_{\mu}$ from \eqref{quad inv a} as
\be
    a_{m+2}[A_{\mu}] = -\frac 12 \int \frac{d^{d}x}{(4\,\pi)^{\frac d2}}\,
    F^{\mu\nu}\,\frac{\left(\frac12\,\partial_{x}^{2}\right)^{\!m}}{(2m+3)!!}
    \,F_{\mu\nu}+\mathcal{O}(A^{3})
    \qquad [m\geqslant0]\,,
\ee where gauge invariance is manifest. The other coefficients
$a_{n\leqslant 1}$ can be also obtained from \eqref{quad non inv a}, but
they are all vanishing except $a_{0}$ which gives a constant. In
particular one can see for $a_{1}$ a cancellation between the
linear part $h^{\sst (0)}_{A_{\mu}}$ and the quadratic part
${(h^{\sst (1)}_{A_{\mu}})}^{\nu}{(h^{\sst
(1)}_{A_{\mu}})}_{\nu}\,/4$\,.

\subsection{Gravitational interaction}

We consider the action of a scalar field $\chi$ in a curved space $g_{\mu\nu}$
with a scalar curvature coupling:
\be
    \frac12\int d^dx\ \sqrt{g} \left(g^{\mu\nu}\partial_{\mu}\chi\,\partial_{\nu}\chi+
    \xi\,\mathcal{R}\,\chi^{2}+m^2\,\chi^2\right).
\ee
In order to rewrite the above as \eqref{S H},
we should first redefine the scalar field as
$\phi_{\chi,g_{\mu\nu}}=g^{\frac{1}{4}}\,\chi/2$ with
$(\phi_{\chi,g_{\mu\nu}})^{*}=\phi_{\chi,g_{\mu\nu}}$, then the action can be written with
\be
    \hat H_{g_{\mu\nu}}=-\left(\hat P_{\mu}-i\,\frac{\partial_{\mu}
    g(\hat X)}{4\,g(\hat X)}\right)
    g^{\mu\nu}(\hat X) \left(\hat P_{\nu}+i\,\frac{\partial_{\nu}g(\hat X)}
    {4\,g(\hat X)}\right)
    -\xi\,\mathcal{R}(\hat X)\,.
\ee Again from the Wigner map, we identify the components of
$h_{g_{\mu\nu}}(x,u)$. First the spin $2$ part is given by the
perturbation of the inverse metric tensor $\kappa^{\mu\nu}$\,: \be
    h^{\sst (2)}_{g_{\mu\nu}}(x,u) = u^{2}-u_{\mu}\,u_{\nu}\,g^{\mu\nu}
    =: u_{\mu}\,u_{\nu}\,\kappa^{\mu\nu}(x)\,,
\ee
and the components 
$h^{\sst (1)}_{g_{\mu\nu}}$ and $h^{\sst (n\geqslant 3)}_{g_{\mu\nu}}$
vanish while the
scalar part $h^{\sst (0)}_{g_{\mu\nu}}$ is given by
\ba
    h^{\sst (0)}_{g_{\mu\nu}}(x)\e -
    \xi\,\mathcal{R}(x)-\frac{g^{\mu\nu}(x)\,\partial_{\mu}g(x)\,
    \partial_{\nu}g(x)}{16\,g^2(x)}
    -\frac{1}{4}\,\partial_{\mu} \partial_{\nu}g^{\mu\nu}(x)-
    \partial_{\mu} \left(
     \frac{g^{\mu\nu}(x)\,\partial_{\nu}g(x)}{4\,g(x)}\right) \nn
     \e -\xi\,\mathcal{R}(x)
    +\frac{1}{4} \left(\partial_{\mu}\partial_{\nu}\kappa^{\mu\nu}
    -\partial^2\kappa' \right)-\frac{1}{16}
    \partial_{\mu}\kappa'\partial^{\mu}\kappa'\nn
    && \quad +\,\frac{1}{4}\,\partial_{\mu}
    \left[(\kappa^{\mu\nu}+\eta^{\mu\nu}\kappa')
    \partial_{\nu}\kappa'\right]
    -\frac{1}{2}\,\partial^2\left[\kappa^{\mu\nu}
    \kappa^{\mu\nu}+(\kappa')^2\right]
    +\mathcal{O}(\kappa^{3})\,.
 \ea
 The non-linear $\epsilon$-transformation \eqref{nonlinear e
transf} defines the general coordinate transformation,  and the
non-linear $\alpha$-transformation \eqref{nonlinear a transf}
for massless ($m=0$) and conformally coupled ($\xi=\xi_{\rm conf}$)
case gives a  Weyl transformation.

 The HKE coefficients $a_{n\geqslant2}$ can be obtained up to the quadratic order
 in $\kappa_{\mu\nu}$  from \eqref{quad inv a} as
 \be
    a_{m+2}[g_{\mu\nu}]=\frac18\int \frac{d^{d}x}{(4\pi)^{\frac d2}}\,
    R_{\mu\nu\rho\sigma}\,\frac{(\frac12\partial_{x}^{2})^{m}}{(2m+5)!!}\,
    R^{\mu\nu\rho\sigma}
    +\frac{c(m,\xi)}2\,R\,\,\frac{(\frac12\partial_{x}^{2})^{m}}{(2m+5)!!}\,R
    +\mathcal{O}(\kappa^{3})\,,
    \label{heat coeff gravity}
\ee
where $c(m,\xi)$ is a constant depending on the HKE coefficient index
and the parameter $\xi$\,:
\be
    c(m,\xi)=(2m+3)(2m+5)(4\xi-1)^{2}
    +2(2m+5)(4\xi-1)+1\,,
\ee 
and $R_{\mu\nu\rho\sigma}$ is 
the linearized Riemann tensor:
\be
    R_{\mu\nu\rho\sigma}=\frac12\left(
    \partial_{\mu}\partial_{\sigma}\kappa_{\rho\nu}
    +\partial_{\rho}\partial_{\nu}\kappa_{\mu\sigma}
    -\partial_{\mu}\partial_{\rho}\kappa_{\nu\sigma}
    -\partial_{\nu}\partial_{\sigma}\kappa_{\mu\rho}\right),
\ee
and $R_{\mu\nu}:=\eta^{\rho\sigma}\,R_{\mu\rho\nu\sigma}$
and $R:=\eta^{\mu\nu}\,R_{\mu\nu}$ are respectively 
the linearized Ricci tensor and the linearized Ricci scalar.
The expression \eqref{heat coeff gravity} 
was chosen by using the following (Gauss-Bonet like) identity:
\be
 	R_{\mu\nu\rho\sigma}\,\Box^{m}\,R^{\mu\nu\rho\sigma}
	-4\,R_{\mu\nu}\,\Box^{m}\,R^{\mu\nu}
	+R\,\Box^{m}\,R=
	(\rm{total\ derivative})\,,
\ee
in such a way that the linearized Ricci tensor
does not appear.

Notice that   when $\xi=\xi_{\rm conf}$\,, the anomaly term
$a_{d/2}$ is given by the square of the Weyl tensor. From
\eqref{quad non inv a}, one can check that $a_{n}$ with negative
$n$ again vanish, and the non-vanishing coefficients $a_{1}$ and
$a_{0}$ are given up to quadratic order in $\kappa_{\mu\nu}$ by
\ba
    a_{1}[g_{\mu\nu}]\e\int \frac{d^{d}x}{(4\,\pi)^{\frac d2}}\,
    \bigg[-\xi \left(1+\frac12\,\kappa'\right) \mathcal{R}+ \nn
    && +\,
    \frac16 \left(\frac14\,(\partial_{u}\kappa')^{2}
    -\frac14\,(\partial_{\rho}\kappa_{\mu\nu})^{2}
    +\frac12\,(\partial^{\nu}\kappa_{\mu\nu})^{2}
    -\frac12\,\partial_{\mu}\kappa'\,\partial_{\nu}\kappa^{\mu\nu}
    \right)+\mathcal{O}(\kappa^{3})\bigg]\,, \nn
    a_{0}[g_{\mu\nu}] \e \int \frac{d^{d}x}{(4\,\pi)^{\frac d2}}\,
    \left[1+\frac12\,\kappa'+\frac18\,\kappa'\,\kappa'
    +\frac14\,\kappa_{\mu\nu}\,\kappa^{\mu\nu}+
    \mathcal{O}(\kappa^{3})\right].
\ea
From the linearized expressions of the Hilbert-Einstein action
and Cosmological constant:
\ba
    && \sqrt{g}\, \mathcal{R} =
    \frac14\,(\partial_{u}\kappa')^{2}
    -\frac14\,(\partial_{\rho}\kappa_{\mu\nu})^{2}
    +\frac12\,(\partial^{\nu}\kappa_{\mu\nu})^{2}
    +({\rm total\ derivative})+\mathcal{O}(\kappa^{3})\,, \nn
    && \sqrt{g}=1+\frac12\,\kappa'+\frac18\,\kappa'\,\kappa'
    +\frac14\,\kappa_{\mu\nu}\,\kappa^{\mu\nu}+
    \mathcal{O}(\kappa^{3})\,,
\ea we can see that $a_{1}$ and $a_{0}$ coincide with
 \be
    a_{1}[g_{\mu\nu}]=\int \frac{d^{d}x}{(4\,\pi)^{\frac d2}}
    \left(\frac16-\xi\right) \sqrt{g}\,\mathcal{R}\,,
    \qquad
    a_{0}[g_{\mu\nu}]=\int \frac{d^{d}x}{(4\,\pi)^{\frac d2}} \sqrt{g}\,,
\ee computed using different methods.

\section{Conclusion}
\label{sec:conclu}

In this work, we considered the most general quadratic action for a scalar field.
We argued that it describes the interaction of the scalar field with a
 background of symmetric tensor fields of arbitrary rank. We gave the perturbative expansion of the quantum
effective action in powers of the ultraviolet cut-off and of the external fields.

Following the \emph{induced gravity} idea of Sakharov (see \textit{e.g.} \cite{Sakharov:1967pk,Visser:2002ew} for a review), one may
consider the cut-off dependent local terms as providing the
dynamics of the tensor fields. This interpretation, if
consistent, would result in an interacting higher-spin gravity theory. 
To prove the consistency one
has to overcome several obstacles. The first one is that
Fronsdal's free action is not recovered. As noted in the
introduction, this can be traced to the presence of linear terms in
the effective action and to their mixing with quadratic terms in
the fields and their derivatives. This raises the questions of 
the existence of a stable vacuum and of the
spectrum of the theory around flat spacetime. In fact, here the gauge fields
and parameters are not subject to any trace constraint and thus do not fit into Fronsdal's
formulation  \cite{Fronsdal:1978rb} so the naive counting of modes can be misleading. 
Moreover, imposing
from the beginning trace constraints on the higher-spin fields
would break the non-Abelian symmetry group of unitary operators. The unconstrained formulation \cite{Francia:2002aa
} and the
triplets \cite{Francia:2002pt
} both rely on unconstrained higher-spin tensors but the local actions involve additional fields as well; the elimination of the latter
typically leads to non-local actions.
Once the issue of the spectrum is
understood, another problem is the infinite number of divergent
terms which appear in the effective action in contrast to the
gravitational case where only a finite number of terms is present.
A resummation of these terms may be necessary. We hope to come
back to this tantalizing \emph{induced higher-spin gravity} interpretation of
the effective action in the near future.

Another important target of applications from our results is the AdS/CFT
correspondence. The operator {$\hat H$} represents the
boundary data for the higher-spin gauge fields on $AdS_{d+1}$.
The matching of the {physical} degrees of freedom obtained from the rank-$s$ external field $\bar h^{\sst(s)}$ with local symmetries \eqref{linear transf} with an on-shell spin-$s$ gauge field in $AdS_{d+1}$ can be seen as follows.\footnote{In the literature, this matching is usually performed for the boundary currents and the gauge-fixed bulk fields, see \textit{e.g.} \cite{Mikhailov:2002bp}.}
A Fronsdal spin-$s$ gauge field on $AdS_{d+1}$ is described by an $O(d,2)$-
covariant rank-$s$ tensor which is
doubly-traceless \cite{Fronsdal:1978vb}, thus it can be decomposed under  the boundary
$O(d,1)$   into four symmetric
tensors of rank $s, s-1, s-2$ and $s-3$. In the same way, the
{corresponding Fronsdal gauge parameter is an $O(d,2)$-covariant rank-$(s-1)$ tensor which}
can be decomposed into
two boundary symmetric tensors of rank $s-1$ and $s-2$. Let
$d_{r}$ {denote\footnote{Although the precise value of this number is not necessary in the present argument, we remind the reader that $d_r=(d+r-1)!/r!(d-1)!$} the number of components of a boundary symmetric
tensor of rank $r$. The number of physical degrees of freedom of
the spin-$s$ \emph{on-shell bulk}} gauge field is equal to
\be
(d_{s}+d_{s-1}+d_{s-2}+d_{s-3})-2\,(d_{s-1}+d_{s-2})\,,
\ee
where we removed twice the number of components of the gauge parameters since the field
is on-shell. This matches the number of physical degrees of freedom of a spin-$s$ \emph{off-shell boundary}
field $\bar h^{\sst (s)}$ with
gauge symmetries \eqref{linear transf}\,: \be
    d_{s}-(d_{s-1}+d_{s-2}-d_{s-3})\,,
\ee where $d_{s-3}$ corresponds to the number of overlapping degrees of freedom between
$\bar\epsilon^{\sst (s-1)}$ and $\bar\alpha^{\sst(s-2)}$. The
number of independent gauge parameters were removed just once since we are
considering an off-shell counting.

This simple counting argument should be extended to a non-trivial match between the Lorentz-covariant quadratic terms in the effective action computed here and the on-shell bulk Fronsdal actions with appropriate boundary counter-terms. This computation would be also of interest in order to compare directly the gauge symmetries on the bulk/boundary sides, as well the anomalies arising from the IR/UV regularizations. Indeed, the full
symmetry group of the boundary data at quantum level has
been identified with
the group of unitary operators. Comparison with the
Vasiliev gauge group is not straightforward because our gauge fields are
metric-like and not frame-like. We leave all these issues for a future work.

\acknowledgments{
We are grateful to
N. Boulanger,
A. Campoleoni,
D. Francia,
C. Iazeolla,
D. Nesterov,
F. Nitti,
R. Rahman,
A. Sagnotti,
S. Solodukhin,
P. Sundell
and
M. Taronna
for many helpful discussions,
and to APC-Paris VII and Scuola Normale Superiore
for the kind hospitality extended to one or more of us.
The present research was supported in part by Scuola Normale Superiore, by INFN and
by the ERC Advanced Investigator Grant no. 226455 ``Supersymmetry, Quantum Gravity and Gauge Fields'' (SUPERFIELDS).
}

\appendix

\section{Notations and useful formulas}
\label{sec:not}

\subsection{Generating functions}\label{genfct}

Symmetric tensors $f^{\sst(s)}_{\mu_1\cdots\mu_s}$ of different rank $s$ are
contracted with auxiliary variable $u^\mu$ to form a \emph{generating function} $f(x,u)$ of these tensors:
\be
    f(x,u):= \sum_{s=0}^\infty f^{\sst (s)}(x,u)\,,\qquad
    f^{\sst(s)}(x,u):= \frac{1}{s!}\,
    f^{\sst(s)}_{\mu_1\cdots\mu_s}(x)\,u^{\mu_1}\cdots u^{\mu_s}\,,
\ee
where the $s$-th order term $u$ is written with the superscript $(s)$.
The contraction $\pd{f(x)}{g(x)}$ of two generating functions $f$ and $g$ is
defined by the sum of the contractions of all of their tensors:
\be
    \pd{f(x)}{g(x)}
    := \sum_{s=0}^\infty \frac{1}{s!}\,f^{\sst(s)}_{\mu_1\cdots\mu_s}(x)\,
    g^{\sst(s)}{}^{\mu_{1}\cdots\mu_{s}}(x)\,,
    \label{inproduct}
\ee
and we define in addition a scalar product by
\be
    \ppd{f}{g} := \int d^dx\ \pd{f(x)}{g(x)}\,.
\ee

\subsection{Higher-spin curvatures}

The spin-$s$ curvature of Weinberg \cite{Weinberg:1964ew} is a mixed-symmetry tensor in the antisymmetric convention (generalizing thereby the linearized Riemann tensor of $s=2$)
and is defined with the aid of the curl operator
$\pi_{\mu\nu}$\,: \be \pi_{\mu\nu}={\partial \over \partial x^\mu}
{\partial \over \partial u^\nu}-{\partial \over \partial x^\nu}
{\partial \over \partial u^\mu}\,,\ee
 as \be
    R^{\sst (s)}_{\mu_{1}\nu_{1} \cdots \mu_{s}\nu_{s}}(x)
    :=\pi_{\mu_1\nu_1}\,\cdots\, \pi_{\mu_s\nu_s}\,h^{\sst (s)}(x,u)=
    \partial^{\,}_{[\mu_{s}}\cdots \partial^{\,}_{[\mu_{1}}\,
    h^{\sst (s)}_{\nu_{1}]\cdots \nu_{s}]}(x)\,,
\ee
such that $R^{\sst(s)}$ is invariant under the gauge transformation,
$\delta\,h_{\mu_{1}\cdots \mu_{s}}=
\partial_{(\mu_{s}} \varepsilon_{\mu_{1}\cdots\mu_{s-1})}$.
In the text, we have used the spin-$s$ curvature of deWit and Freedman \cite{deWit:1979pe} which is a mixed-symmetry tensor in the symmetric convention (generalizing thereby the linearized Jacobi tensor of $s=2$) defined via two auxiliary variables as   
\ba
    R^{\sst (s)}(x,v,w) &:=&
    \frac1{s!}\,
    R^{\sst(s)}_{\mu_{1}\nu_{1} \cdots \mu_{s}\nu_{s}}(x)\,
    v^{\mu_1}\cdots v^{\mu_s}\,w^{\nu_1}\cdots w^{\nu_s}\nn
    &\ =& \frac1{s!}\,\big(v\cdot\partial_x\,w\cdot\partial_u
    -w\cdot\partial_x\,v\cdot\partial_u\big)^s\,h^{\sst (s)}(x,u)\,,\\
    R(x,v,w) &:=&
    \sum_{s=0}^{\infty} R^{\sst (s)}(x,v,w) =
    e^{v\cdot\partial_{x}\,w\cdot \partial_{u}-w\cdot\partial_{x}\,v\cdot \partial_{u}}\,
    h(x,u)\,\Big|_{u=0} \,.
    \label{HS curvature}
\ea
It enjoys the following properties:
\ba
    && R^{\sst (s)}(x,w,v) = (-1)^{s}\,R^{\sst (s)}(x,v,w)\,, \nn
    && w\cdot \partial_{v}\,R(x,v,w)=v\cdot \partial_{w}\,R(x,v,w)=0\,.
\ea

\subsection{Weyl/Wigner quantization}
\label{sec:Weyl}

The Weyl/Wigner quantization \cite{Weyl:1927vd
} is a method for systematically associating a
(pseudo)differential operator with a distribution in phase space.
It offers a classical-like formulation of quantum
mechanics using real functions on phase space 
as observables and the Wigner function as an
analogue of the Liouville density function.

The \emph{Weyl map} associates to a distribution $f$
a Weyl(\textit{i.e.} symmetric)-ordered operator $\hat{F}$ defined by
\be
    \hat{F} =\int \frac{d^d k\,d^d y}{(2\,\pi)^{d}}
    \,\,{\cal F}(k,y)\,
    e^{i\,( k\cdot \hat{X}-y\cdot\hat{P})}\,,
    \label{Weylmap}
\ee
where ${\cal F}$ is the Fourier transform of $f$ over
the whole phase space:
\be
    {\cal F}(k,y) = \int \frac{d^dx\,d^dp}{(2\pi)^{d}}\,
    f(x,p)\,e^{-i\,(k\cdot x - y\cdot\,p)}\,,
\ee
and the function $f(x,p)$ is called the \emph{Weyl symbol} of the operator $\hat{F}\,$.
The inverse of the Weyl map is called the \emph{Wigner map}:
\be
    f(x,p) = \int d^dy\,\,\langle\,x-y/2\mid\hat{F}\mid
    x+y/2\,\rangle\,\,e^{i\, y \cdot p}\,.
    \label{Wignermap}
\ee
A nice property of the these maps is that it relates
the complex conjugation $^*$ of symbols to the Hermitian
conjugation $^\dagger$ of operators. Consequently, the
image of a real function is a Hermitian operator.

The \emph{Moyal product} $\star$ is the pull-back of the
composition product in the algebra of quantum observables with
respect to the Weyl map, such that the latter becomes
an isomorphism of associative algebras.
The Wigner map \eqref{Wignermap} allows to check that the
following explicit expression of the Moyal product:
\be
    f(x,p) \star g(x,p)=
    e^{\frac i2\,(\partial_{x_{1}}\!\cdot\, \partial_{p_{2}}
    -\partial_{x_{2}}\!\cdot\, \partial_{p_{1}})}\,
    f(x_{1},p_{1})\,g(x_{2},p_{2})\,
    \Big|_{\overset{x_{1}=x_{2}=x}{\sst p_{1}=p_{2}=p}}\,.
\ee

The trace of an operator is given by integral of its Weyl symbol
over phase space:
\be
    \mbox{Tr}[\,\hat{F}\,] =
    \int \frac{d^{d}x\,d^{d}p}{(2\,\pi)^{d}}\  f(x,p)\,,
\ee
and the trace formula for a product of operators leads to
\ba
    \mbox{Tr}[\,\hat{F}\,\hat{G}\,] \e
    \int \frac{d^{d}x\,d^{d}p}{(2\,\pi)^{d}}\  f(x,p) \star g(x,p)=
    \int \frac{d^{d}x\,d^{d}p}{(2\,\pi)^{d}}\  f(x,p) \, g(x,p)\,.
    \label{star trace}
\ea

\subsection{Special functions}

Several useful definitions and formulas for Bessel function
are collected here in order to be self-contained.

The \emph{Bessel function of first kind} can be defined as a series by
\be
    J_{\nu}(z):=\sum_{m=0}^{\infty}
    \frac{(-1)^{m}}{m!\,\Gamma(\nu+m+1)} \left(\frac z2\right)^{2m+\nu}\,,
\ee
and is related to the \emph{modified Bessel function} $I_{\nu}$ by
\be
    I_{\nu}(z):=i^{-\nu}\,J_{\nu}(i\,z)\,.
\ee
For integer order $\nu=n$, $J_{n}$ and $I_{n}$ can be generated as
\be
    e^{\frac z2\, (t-\frac1t)}=\sum_{n=-\infty}^{\infty}\,J_{n}(z)\,t^{n}\,,
    \qquad
    e^{\frac z2\, (t+\frac1t)}=\sum_{n=-\infty}^{\infty}\,I_{n}(z)\,t^{n}\,.
    \label{bessel gen}
\ee
The Bessel function $J_{\nu}$ admits also the following integration identity:
\ba
    && \int_{0}^{\frac\pi2}d\theta\,\sin^{\mu+1}\theta\,
    J_{\mu}(z_{1}\,\sin\theta)\,
     \cos(z_{2}\,\cos\theta)= \nn
     &&\quad =\,\sqrt{\frac\pi2}\,z_{1}^{\mu}\,
     \left(\sqrt{z_{1}^{2}+z_{2}^{2}}\right)^{\!\!-(\mu+\frac12)}
     J_{\mu+\frac12}\!\left(\sqrt{z_{1}^{2}+z_{2}^{2}}\right)
     \qquad
     [\mathrm{Re}(\mu)>-1]\,.
     \label{int. Bessel}
\ea
The Lommel's expansion of the Bessel function is given by
\be
    (z+\omega)^{-\frac\nu2}\,J_{\nu}(\sqrt{z+\omega})
    =\sum_{k=0}\,\frac{(-\frac12\,\omega)^{k}}{k!}\,
    z^{-\frac{\mu+k}2}\,J_{\nu+k}(\sqrt{z})\,,
    \label{Lommel}
\ee
and the addition theorem gives
\be
    Z^{-\nu}\,J_{\nu}(Z)=
    2^{\nu}\,\Gamma(\nu)\,\sum_{m=0}^{\infty}\,(\nu+m)\,
    z_{1}^{-\nu}\,J_{\nu+m}(z_{1})\,z_{2}^{-\nu}\,J_{\nu+m}(z_{2})\,C^{\nu}_{m}(\cos\theta)\,,
\ee
where $Z^{2}=z_{1}^{2}+z_{2}^{2}+2\,z_{1}\,z_{2}\,\cos\theta$
and $C^{\nu}_{m}$ is the \emph{Gegenbauer polynomial} defined by
\be
    C^{\nu}_{m}(z)=
    \sum_{k=0}^{[\frac m2]}
    \frac{(-1)^{k}\,(\nu)_{m-k}}{m!\,(m-2k)!}\,(2\,z)^{m-2k}
    \qquad [\nu>-{\st \frac12},\, \nu\neq0 ]\,.
\ee
The Gegenbauer polynomial satisfies the following differential equation:
\be
    \left[ \left(1-z^{2}\right) \frac{d^{2}\ }{dz^{2}}
    -(2\,\nu+1)\,z\,\frac{d\ }{dz} + m(m+2\,\nu)\right]
    C^{\nu}_{m}(z)=0\,.
    \label{de Gen}
\ee

\section{Computational appendices}
\label{sec:app comp}

\subsection{Inverse map of the projection-like operator}
\label{sec:inverse}

To check whether $\Pi^{-1}_{d}(q,\partial_{x})$ \eqref{Pi'} is really the inverse of
$\Pi_{d}(q,\partial_{x})$ \eqref{Pi}, we compute directly
 $ \Pi^{-1}_{d}(q,\partial_{x})\,\Pi_{d}(q,\partial_{x})$\,:
\ba
    \Pi^{-1}_{d}(q,\partial_{x})\,\Pi_{d}(q,\partial_{x}) \e
    \sum_{m,n=0}^{\infty}\, \frac1{(q\cdot\partial_{q}+\frac{d-1}2-2m)_{m}\,
    (-q\cdot\partial_{q}-\frac{d-5}2+2m)_{n}}\,
    \frac{\left(\frac14\,P(q,\partial_{x})\right)^{m+n}}{m!\,n!} \nn
    \e \sum_{N=0}^{\infty}\,\frac{c_{N}(q\cdot\partial_{q}+{\frac{d-3}2})}{N!}
    \left(\frac14\,P(q,\partial_{x})\right)^{N}\,,
\ea
and $\Pi_{d}(q,\partial_{x})\,\Pi^{-1}_{d}(q,\partial_{x})$\,:
\ba
    \Pi_{d}(q,\partial_{x})\,\Pi^{-1}_{d}(q,\partial_{x})\e
    \sum_{m,n=0}^{\infty}\,\frac1{(-q\cdot\partial_{q}-\frac{d-5}2)_{m}\,
    (q\cdot\partial_{q}+\frac{d-1}2-2N)_{n}}\,
    \frac{\left(\frac14\,P(q,\partial_{x})\right)^{m+n}}{m!\,n!} \nn
    \e \sum_{N=0}^{\infty}\,\frac{d_{N}(q\cdot\partial_{q}+{\st \frac{d-3}2})}{N!}
    \left(\frac14\,P(q,\partial_{x})\right)^{N}\,,
\ea
where we collected terms of the same order in $P$
with coefficients $c_{N}(x)$ and $d_{N}(x)$ as series:
\ba
    c_{N}(x) \e \sum_{n=0}^{N}\, \binom{N}{n}\,
    \frac1{(x-2n+1)_{n}\,(-x+2n+1)_{N-n}}\,,\nn
    d_{N}(x) \e \sum_{n=0}^{N}\,\binom{N}{n}\,
    \frac1{(x+1)_{n}\,(-x+2N+1)_{N-n}}\,.
\ea
Finally one can show with a help of Mathematica that
\be
    c_{N}(x)=d_{N}(x)=\delta_{N,0}\,.
\ee

\subsection{Expression by higher-spin curvature tensors}
\label{sec:curvature} 

Here we prove the equation \be
f([\partial_{v}\partial_{w}])\,
    e^{\frac12 \left<\, [vw]\,[xy]\,\right>}\,\Big|_{v=w=0}
    =\int_{0}^{\infty} dt\,t\,e^{-t}\,f(-t\,[xy])\,,
\label{equat} \ee which was used in equation \eqref{transf R}.
First write $f([\partial_v\partial_w])$ as
 \be
    f([\partial_v\partial_w])=
    \int d\mu(A)\,\tilde f(A)\,e^{\frac i2 \left<\,A\,[\partial_v\partial_w]\,\right>}.
\ee Then  use the notation $V:=v\oplus w$ and define the two
$2d\times 2d$   matrices: \be
    B=i\left(\begin{array}{cc}
    0 & A \\ -A & 0
    \end{array}\right)\,,
    \qquad
    S=\left(\begin{array}{cc}
    0 & [xy] \\ -[xy] & 0
    \end{array}\right)\,,
\ee
so that  \be
    i\left<A\,[\partial_{v}\partial_{w}]\,\right>
    ={\partial_{V}}^{t}\,B\,\partial_{V}\,,
    \qquad
    \left<\,[vw]\,[xy]\,\right>=V^{t}\,S\,V\,.
\ee Next, we express the above expression using a Gaussian
integral as\ba
    &&e^{\frac i2 \left<\,A\,[\partial_v\partial_w]\,\right>}
    \,e^{\frac 12 \left<\,[vw]\,[xy]\,\right>}\,\Big|_{v=w=0}
    =e^{\frac12\, {\partial_{V}}^{t}\,B\,\partial_{V}}\,
    e^{\frac12\,V^{t}\,S\,V}\,\Big|_{V=0} \nn
    &&= ( \det B )^{-\frac12} \int \frac{d^{2d}X}{\pi^{d}}
    \,e^{-\frac12\,X^{t}\,B^{-1}\,X+X\cdot \partial_V}\,
     ( \det S )^{-\frac12} \int \frac{d^{2d}Y}{\pi^{d}}
    \,e^{-\frac12\,Y^{t}\,S^{-1}\,Y+Y\cdot V} \nn
    &&= \det(1-S\,B)^{-\frac12} = \det (1+ i\,[xy]\,A)^{-1}
    = e^{-\left<\,\ln(1+i\,A\,[xy]\,)\right>}\,.
\ea By making use of  \be
    \left<\,(A\,[xy]\,)^{n}\,\right>
    =2\left[\frac 12\,\left< A\,[xy]\,\right>\right]^{n}\,,
\ee we get \be
    e^{\frac i2 \left<\,A\,[\partial_v\partial_w]\,\right>}
    \,e^{\frac 12 \left<\,[vw]\,[xy]\,\right>}\,\Big|_{v=w=0}
    = \frac1{\left(1+\frac i2\left< A\,[xy]\,\right>\right)^{2}}
    =\int_{0}^{\infty} dt\, t\, e^{-t \left(1+\frac i2 \left< A\,[xy]\,\right> \right)}\,,
\ee and finally we obtain the  the desired relation \eqref{equat}.

\subsection{Finite part of the effective action}
\label{sec:int. 1F1}

To compute the effective action, we now need to evaluate the
following definite integral of hypergeometric function:
\be
    \int_{\varepsilon^{2}}^{\infty} dt\ t^{-\frac{d-4}2-1}\,
    {}_{1}F_{1}\!\left(1;m+\frac32;-\,\frac{p^{2}}4\,t\right)\,.
    \label{definite}
\ee
To do so, let us first consider the indefinite integral:
\ba
    && \int  dt\ t^{-\nu-1}\,
    {}_{1}F_{1}\!\left(1;m+\frac32;-\,\frac{p^{2}}4\,t\right)=
    \sum_{n=0}^{\infty}\,\frac{(-\frac{p^{2}}4)^{n}}{(m+\frac32)_{n}}\,
    \frac{t^{-\nu+n}}{-\nu+n}
    \nn
    && \quad\,
    =-\,\frac{t^{-\nu}}{\nu}\,_{2}F_{2}\!\left(
    1,-\nu;m+\frac32,1-\nu;-\,\frac{p^{2}}4\,t\right)
    \qquad [\,\nu\notin \mathbb{N}\,]\,,
    \label{indefinite}
\ea
where we evaluated the integral by using the series representation
of hypergeometric function $_{1}F_{1}$ and
re-expressed the result by another hypergeometric function $_{2}F_{2}$.
To compute the definite integral \eqref{definite}, we need to determine
the integration constant which is given by
\ba
    &&\lim_{t\to\infty}\,-\,\frac{t^{-\nu}}{\nu}\,_{2}F_{2}\!\left(
    1,-\nu;m+\frac32,1-\nu;-\,\frac{p^{2}}4\,t\right)= \nn
    && \qquad = -\,\frac{\pi}{\sin(\pi\,\nu)}\,
    \frac{\Gamma(m+\frac32)}{\Gamma(m+\frac32+\nu)}\,
    \left(\frac{p^{2}}4\right)^{\nu}
    \qquad [\,p^{2}\geqslant0,\,\nu>-1,\,\nu\notin \mathbb{N}\,]\,.
    \label{int const}
\ea
For odd $d$, we obtain \eqref{definite} as
\be
     \pi\,i\,\frac{ \left(-\,\frac{p^{2}}4\right)^{\frac{d-4}2}}{(m+\frac32)_{\frac{d-4}2}}
     +
    \varepsilon^{-d+4}\,
    \sum_{n=0}^{\infty}\,
    \frac{\left(-\,\frac{p^{2}}4\,\varepsilon^{2}\right)^{n}}
    {(m+\frac32)_{n}\,(\frac{d-4}2-n)}\,.
\ee
For even $d$, we consider $\nu=(d-4)/2+\xi$ and take
$\xi\to 0$ limit. The pole $\xi^{-1}$ from the
indefinite integral \eqref{indefinite} cancels out
the pole from the integration constant \eqref{int const},
and we can take the limit smoothly and obtain \eqref{definite} as
\be
    -\,\frac{\left(-\,\frac{p^{2}}4\right)^{\frac{d-4}2} }{(m+\frac32)_{\frac{d-4}2}}
    \left[ \psi\!\left(m+\frac{d-1}2\right)+
    \ln\!\left(\frac{p^{2}}4\,\varepsilon^{2}\right)\right]
    +\ \varepsilon^{-d+4}\hspace{-12pt}
    \sum_{n\neq \frac{d-4}2, n=0}^{\infty}\,
    \frac{\left(-\,\frac{p^{2}}4\,\varepsilon^{2}\right)^{n}}
    {(m+\frac32)_{n}\,(\frac{d-4}2-n)}\,.
\ee

\subsection{Two-point correlation function}
\label{sec:correlation}
The generating function of two-point function
can be expanded by using successively the Newton's binomial series as
\ba
    && \frac1{\left[ (x^{2}+q^{2})^{2}-4\,(x\cdot q)^{2}\right]^{\frac{d-2}2}}
    = \frac1{(x^{2}+q^{2})^{d-2}}
    \sum_{m=0}^{\infty} \binom{m+\frac{d-4}2}{m}
    \left(4\,\frac{(x\cdot q)^{2}}{(x^{2}+q^{2})^{2}}\right)^{m}\nn
    &&\qquad\,=\frac1{\left(x^{2}\right)^{d-2}}
    \sum_{m=0}^{\infty}\left(-\frac{(x\cdot q)^{2}}{x^{2}\,q^{2}}\right)^{m}
    \sum_{n=m}^{\infty}
    \left(-\frac{q^{2}}{x^{2}}\right)^{n}\,
     b_{m,n}\,,
     \label{2corr exp}
\ea
with
\be
    b_{m,n}=    4^{m}\,\binom{m+\frac{d-4}2}{m}
    \binom{n+m+d-3}{n-m}\,.
\ee
On the other hand, one can show, by induction, how
$P(q,\partial_{x}):= [(q\cdot\partial_{x})^{2}-q^{2}\,\partial_{x}^{2}]/4$
acts on $(x^{2})^{-d+2}$:
\be
    P^{n}(q,\partial_{x})\,\frac{1}{\left(x^{2}\right)^{d-2}}
    =\sum_{m=0}^{n}\,\binom{n}{m}\,
    (d-2)_{n+m}\,\left(\frac{d-1}2+m\right)_{n-m}
    \frac{(x\cdot q)^{2m} \left(-q^{2}\right)^{n-m}}
    {\left(x^{2}\right)^{d-2+n+m}}\,.
\ee
Then an arbitrary series of $P(q,\partial_{x})$ acts on $(x^{2})^{-d+2}$ as
\ba
    \left(\sum_{n=0}^{\infty}\,a_{n}\,P^{n}(q,\partial_{x})\right)
    \frac1{\left(x^{2}\right)^{d-2}} \e
    \frac1{\left(x^{2}\right)^{d-2}}
    \sum_{m=0}^{\infty}\left(-\frac{(x\cdot q)^{2}}{x^{2}\,q^{2}}\right)^{m}
    \sum_{n=m}^{\infty} \left(-\frac{q^{2}}{x^{2}}\right)^{n}\times\nn
    && \qquad \times\,
    a_{n}\,\binom{n}{m}\,
    (d-2)_{n+m}\,\left(\frac{d-1}2+m\right)_{n-m}\,,\quad
\ea
and this series coincides to \eqref{2corr exp} with
the choice of $a_{n}$ as
\be
    a_{n}=\frac{1}{n!\,(\frac{d-1}2)_{n}}\,.
\ee
In fact, this coefficient $a_{n}$ gives the Bessel function and
we obtain \eqref{corr Bessel}.

\bibliographystyle{JHEP.bst}

\providecommand{\href}[2]{#2}\begingroup\raggedright\endgroup

\end{document}